\documentclass[epj]{svjour}

\usepackage{amsmath}
\usepackage{graphicx}
\usepackage{epsfig}
\usepackage{float}
\usepackage{amssymb}
\usepackage{color}
\usepackage{subfigure}
\usepackage{sidecap}
\usepackage[normalem]{ulem}  % \sout{old text} for strikeout

\newcommand{\midtilde}{\raisebox{-0.25\baselineskip}{\textasciitilde}}

\begin{document}

\bibliographystyle{ieeetr}

\title{Symmetry energy impact in simulations of core-collapse supernovae}

\titlerunning{Symmetry energy impact in simulations of core-collapse supernovae}

\author{Tobias Fischer\inst{1}, Matthias Hempel\inst{2}, Irina Sagert\inst{3}, Yudai Suwa\inst{4}, and J{\"u}rgen Schaffner-Bielich\inst{5}}

\authorrunning{Fischer et al.}

\institute{
Institute for Theoretical Physics, University of Wroclaw, pl. M. Borna 9, 50-204 Wroclaw, Poland,
\and
Departement Physik, Universit\"at Basel, Klingelbergstrasse 82, CH-4056 Basel, Switzerland,
\and
National Superconducting Cyclotron Laboratory, Michigan State University, 640 S. Shaw Lane, East Lansing, Michigan, US,
\and
Yukawa Institute for Theoretical Physics, Kyoto University, Oiwake-cho, Kitashirakawa, Sakyo-ku, Kyoto 606-8502, Japan,
\and
Institut f\"ur Theoretische Physik, Goethe Universit\"at, Max von Laue Str. 1, D-60438 Frankfurt am Main, Germany
}
\abstract{
We present a review of a broad selection of nuclear matter equations of state (EOSs) applicable in core-collapse supernova studies. The large variety of nuclear matter properties, such as the symmetry energy, which are covered by these EOSs leads to distinct outcomes in supernova simulations. Many of the currently used EOS models can be ruled out by nuclear experiments, nuclear many-body calculations, and observations of neutron stars. In particular the two classical supernova EOS describe neutron matter poorly. Nevertheless, we explore their impact in supernova simulations since they are commonly used in astrophysics. They serve as extremely soft and stiff representative nuclear models. The corresponding supernova simulations represent two extreme cases, e.g., with respect to the protoneutron star (PNS) compactness and shock evolution. Moreover, in multi-dimensional supernova simulations EOS differences have a strong effect on the explosion dynamics. Because of the extreme behaviors of the classical supernova EOSs we also include DD2, a relativistic mean field EOS with density-dependent couplings, which is in satisfactory agreement with many current nuclear and observational constraints. This is the first time that DD2 is applied to supernova simulations and compared with the classical supernova EOS. We find that the overall behaviour of the latter EOS in supernova simulations lies in between the two extreme classical EOSs. As pointed out in previous studies, we confirm the impact of the symmetry energy on the electron fraction. Furthermore, we find that the symmetry energy becomes less important during the post bounce evolution, where conversely the symmetric part of the EOS becomes increasingly dominating, which is related to the high temperatures obtained. Moreover, we study the possible impact of quark matter at high densities and light nuclear clusters at low and intermediate densities.
}
\date{\today}
\PACS{
%04.30.-w Gravitational waves
%04.30.Tv Gravitational-wave astrophysics 
%12.38.-t %Quantum chromodynamics 
%12.38.Mh %Quark-gluon plasma
%21.65.Qr %Quark matter 
%25.75.Nq,
%26.30.Jk 	%Weak interaction and neutrino induced processes, galactic radioactivity 
%26.30.-k %Nucleosynthesis in novae, supernovae, and other explosive environments 
%26.50.+x %Nuclear physics aspects of novae, supernovae, and other explosive environments
26.60.-c %Nuclear matter aspects of neutron stars 
26.60.Kp %Equations of state of neutron-star matter 
%95.85.-e Astronomical observations (additional primary heading(s) must be chosen with these entries to represent the astronomical objects and/or properties studied) 
%95.30.Sf %Relativity and gravitation or 97.60.Lf Black holes,
%95.55.Vj %Neutrino, muon, pion, and other elementary particle detectors
%95.85.Ry %Neutrino, muon, pion, and other elementary particles; cosmic rays 
%95.85.Sz Gravitational radiation, magnetic fields, and other observations 
97.60.Bw %Supernovae
}
\maketitle
%%%%%%%%%%%%%%%%%%%%%%%%%%%%%%%%%%%%%%%%%%%
\section{Introduction}
%TF: supernova phenomenology

Stars more massive than roughly 8 times the mass of our sun (M$_\odot$) end their life as core-collapse supernovae~\cite{Janka:2007,Janka:2012}. These are triggered by the contraction of the stellar core as degenerate electrons are captured on nuclei reducing the main pressure component. When normal nuclear matter density is reached in the very center of the collapsing stellar core, the short-range repulsive force of the strongly interacting nucleon gas counterbalances gravity and the collapse halts. The core bounces back accompanied by the formation of a hydrodynamic shock wave. Initially, the bounce shock breaks out of the high-density core, fully dissociating infalling heavy nuclei into free nucleons and light clusters. The central object that forms at core bounce is the protoneutron star (PNS). It is hot and lepton rich in which sense it differs from the final supernova remnant, the neutron star. The initially expanding shock wave continuously looses energy from the dissociation of heavy nuclei and 
emission of electron-neutrinos when crossing the neutrinospheres. The latter are the spheres of last scattering outside of which neutrinos are freely streaming. The outburst of $\nu_e$, which are produced from local electron captures on free protons, occurs on a short timescale of the order of $5-10\:$ms after core-bounce. Both sources of energy loss, the dissociation of heavy nuclei and neutrino emission, turn the expanding shock wave into a standing accretion front which stalls at $100-150\:$km. The corresponding timescale is about $50-100\:$ms after core-bounce and are given mainly by the progenitor star.

The supernova problem is related to the onset of the explosion in terms of reviving the standing shock wave, i.e. liberating energy from the PNS into the region behind the shock. Note that this relates to the delayed onset of the explosion, which is currently considered to be the standard scenario. Prompt explosions, where the initially expanding bounce shock does not stall, are ruled out. Several explosion scenarios have been explored in the past, the magneto-rotational~\cite{LeBlanc:1970kg}, the acoustic~\cite{Burrows:2005dv}, the high-density quark-hadron phase transition~\cite{Sagert:2008ka}, and the neutrino heating~\cite{Bethe:1985ux} mechanisms. The latter is currently the most favored scenario. However, sophisticated supernova simulations, which include three-flavor Boltzmann neutrino transport and a detailed nuclear equation of state (EOS), obtain explosions in spherical symmetry only for low-mass progenitor stars with $8-9\:$M$_\odot$~\cite{Kitaura:2006,Fischer:2009af}. This is related to the 
special structure of such progenitors. Their low-mass core of about $1.376\:$M$_\odot$ is surrounded by a low-density helium-rich hydrogen envelope, separated by a steep density gradient~\cite{Nomoto:1987,Jones:2013}. This structure leads to an early onset of shock revival via neutrino heating at about 30--40~ms after core bounce. 

More massive stars experience an extended period of mass accretion  which lasts for several $100\:$ms. During this accretion period, the enclosed mass inside the PNS grows and the PNS contracts accordingly. The corresponding timescales are determined by the mass accretion rate which is dependent on the progenitor star and the high-density EOS. In spherically symmetric simulations, neutrino heating is insufficient and fails to revive the standing accretion shock. It requires multi-dimensional simulations where convection and hydrodynamic instabilities increase the neutrino heating  efficiency~\cite{Mueller:2012,Suwa:2013,Bruenn:2013}.

The nuclear symmetry energy, in particular the free symmetry energy associated with the free energy due to the finite and even high temperatures reached, enters supernova simulations via the nuclear matter EOS. In this article we review most of the currently used EOS for supernova matter. Moreover, recent constraints from Chiral Effective Field theory (EFT)~\cite{Hebeler:2010a,Hebeler:2010b,Steiner:2012,Holt:2012a,Sammarruca:2012,Tews:2013,Coraggio:2013} allow us to favor several of these supernova EOSs above others. Unfortunately, the disfavored EOSs include the most commonly used classical EOSs of ref.~\cite{Lattimer:1991nc} (hereafter LS) and of ref.~\cite{Shen:1998gg} (hereafter STOS), despite being consistent with neutron star maximum masses of about $2\:$M$_\odot$. Neutron star radius measurements in low-mass X-ray binaries~\cite{Steiner:2010} could pose tight constraints, indicating that $R_{1.44~\text{M}_{\odot}}=10.4-12.9$~km. However, not yet considered systematic uncertainties may increase the 
error-bars significantly. A compilation of various different probes for the symmetry energy and its slope parameter was recently given in ref.~\cite{Lattimer:2013}, including implications of the two aforementioned constraints.

We apply several supernova EOSs to simulations of stellar collapse and study the resulting SN evolution to identify the impact of the nuclear matter properties and the available experimental and theoretical constraints.  As reference cases, we select the two classical but extreme EOSs LS (a very soft non-relativistic approach) and STOS (a very stiff relativistic-mean field (RMF) approach which utilizes the TM1 interactions \cite{Suga94}). In addition, we apply the RMF approach DD2 with density dependent couplings from ref.~\cite{Typel:2009sy} which matches nuclear constraints at low and intermediate densities, as well as a large neutron star maximum mass~\cite{Lattimer:2013}. Furthermore, this model goes beyond the single nucleus approximation (SNA) employed in the two classical EOSs by including the detailed distribution of several thousands of different nuclei. DD2 has not been used in supernova simulations so far. It is part of the comprehensive supernova-EOS catalogue for the extended nuclear statistical 
equilibrium model of ref.~\cite{Hempel:2009mc} (hereafter HS) which is available online (see below sec.~2.). Several of the EOSs of this catalogue, have already been compared in core-collapse supernova studies~\cite{Hempel:2012,Steiner:2013}. Other recent approaches for the description of supernova matter will be briefly discussed in Sec.~\ref{sec:sneos}.

The appearance of additional degrees of freedom such as hyperons and quarks at supra-saturation densities has long been studied for  cold neutron stars~\cite{Schulze:2011} and during the PNS evolution~\cite{Pons:1999}. At present, little is known about the hyperon-hyperon interactions or many-body forces including hyperons~\cite{Stone:2007,Dexheimer:2010,Bednarek:2012,Weissenborn:2012}. For quark matter, strong QCD interactions have been shown to provide sufficient pressure to support high neutron star masses~\cite{Ozel:2010,Weissenborn:2010,Bonnano:2012,Blaschke:2009,Kurkela:2010,Klaehn:2010,Chen:2011}. From the current understanding of strong interactions, neither hyperon nor quark matter can be ruled out as a component of dense neutron star matter. Depending on the model, they can be both consistent with nuclear physics and give large neutron star masses. Moreover, the large uncertainty in the properties of high-density nuclear matter results in a relatively large freedom in the exploration of the quark or hyperon impact in core-collapse supernova studies~\cite{Takahara:1988yd,Gentile:1993ma,Nakazato:2012,Nakazato:2010,Sumiyoshi:2009,Sagert:2008ka,Fischer:2011,Fischer:2012b,Sagert:2012}. With that, we construct a quark-hadron hybrid EOS (hereafter QB), that allows for large hybrid star maximum masses of $\gtrsim2.01$~M$_\odot$. We apply this quark bag hybrid model in addition to the hadronic EOSs LS220, STOS, and DD2, to simulate the potential impact of quark matter.

The manuscript is organized as follows. In sec.~2, we discuss the supernova matter conditions which must be covered when modeling EOSs applicable for core-collapse supernova studies. We give an overview of the supernova EOS models which are used in the present study. In sec.~3 we discuss their characteristics, such as nuclear matter properties at saturation density, energy per baryon of cold neutron matter, and the neutron-star mass-radius relations and compare these quantities with available constraints. In sec.~4 we explore the impact of the selected EOSs in core-collapse supernova simulations in both spherical symmetry and axial symmetry. Moreover, we discuss the potential impact of light clusters. The manuscript closes with a summary in sec.~5.

\section{Supernova equations of state}

\subsection{General Overview}

\begin{figure}[ht!]
\centering
\includegraphics[width=1.0\columnwidth]{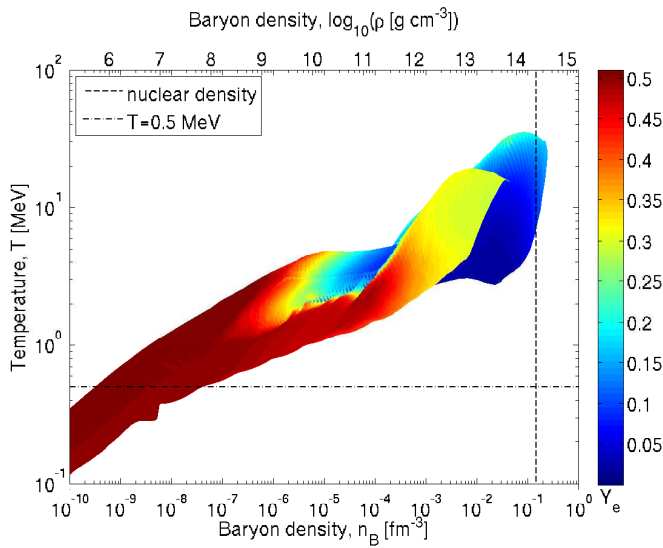}
\caption{Temperature and density (lower scale shows the baryon density and the upper scale shows the restmass density) reached during a standard core-collapse supernova simulation at several $100\:$ms post bounce. The color-coding shows the electron fraction $Y_e$. (color version online)}
\label{fig:phasediagram_sn}
\end{figure}

Fig.~\ref{fig:phasediagram_sn} illustrates the large variety of conditions which has to be handled by a supernova EOS. At temperatures below $\sim0.5$~MeV, time-dependent strong and weak reactions are important to determine the nuclear composition which is dominated by heavy nuclei and is initially given by the progenitor model. In this regime, nuclear $\alpha$-reaction networks are commonly used which include about 14--20 nuclear species. The nuclear EOS has to be able to reproduce the ideal gas of iron-group nuclei which at temperatures of $\sim 0.5\:$MeV reaches a state of chemical and thermal equilibrium, known as nuclear statistical equilibrium (NSE). In NSE, the nuclear EOS can be determined from  three independent variables: the temperature $T$, the rest-mass density $\rho$ (alternatively the baryon number density, $n_B$\footnote{The restmass density used here is related to the baryon number density by $\rho=m_B n_B$, with $m_B=1.674\times 10^{-24}$~g as an (arbitrary) reference mass.}), and the total 
proton-to-baryon ratio $Y_p$ which is equal to the electron faction $Y_e$ due to charge neutrality. Heavy nuclei exist at densities up to normal nuclear matter density $\rho_0 \simeq 2.5\times10^{14}$~g~cm$^{-3}$ ($n_0\simeq0.15$~fm$^{-3}$) and temperatures below $\sim5$~MeV. They are most relevant during the contraction of the stellar core, when the entropy per baryon is low on the order of a few $k_B$. At higher temperatures and densities close to and above $\rho_0$, nuclei dissolve into uniform matter composed of nucleons. The transition region where (heavy) nuclear clusters and free nucleons co-exist is known as inhomogeneous nuclear matter. 

\subsection{Hadronic SN EOS models}
\label{sec:sneos}
The classical supernova EOSs for NSE conditions are those from refs.~\cite{Lattimer:1991nc} (LS) and \cite{Shen:1998gg} (STOS). We apply both in the present study. LS is based on the liquid-drop model and includes surface effects as well as a Maxwell-Boltzmann gas for $\alpha$ particles. It has been provided to users in form of routines for three different values of the compressibility modulus, $180\:$MeV, $220\:$MeV, and $375\:$MeV. STOS is based on the RMF description of homogeneous nuclear matter, combined  with the Thomas-Fermi approach for heavy nuclei and a Maxwell-Boltzmann gas for $\alpha$ particles. Both classical supernova EOSs use the SNA at conditions where heavy nuclei are present (mainly at low entropy per baryon) with an average representative atomic mass and charge. In particular, light nuclear clusters are not considered.

Recently, a new supernova EOS model has been provided which goes beyond the SNA~\cite{Hempel:2009mc} (hereafter HS). It is based on the extended nuclear statistical model of ref.~\cite{Hempel:2009mc} and includes a detailed nuclear composition with up to 10000 nuclear species. This EOS model thereby allows for investigations of additional structures, such as light nuclear clusters at sub-saturation density and their potential impact on supernova dynamics as well as the neutrino signal. For the nucleon interactions, various different RMF models are included. The EOS tables are available online for the parametrizations TM1 \cite{Suga94}, TMA \cite{toki95}, FSUgold \cite{todd2005}, IUFSU \cite{fattoyev2010}, DD2 \cite{Typel:2009sy}, NL3 \cite{Lala97}, SFHo, and SFHx \cite{Steiner:2013} (see Table~\ref{tab:exp_prop1} on the personal homepage of one of the authors\footnote{\label{link_hempel}\texttt{http://phys-merger.physik.unibas.ch/\midtilde hempel/eos.html}}, the 
comprehensive CompOSE EOS database\footnote{\label{link_compose}\texttt{http://compose.obspm.fr}}, 
and the stellarcollapse.org page\footnote{\label{link_ott}\texttt{http://stellarcollapse.org/equationofstate}}). The two new RMF parametrizations SFHo and SFHx have been deduced only very recently~\cite{Steiner:2013} and are motivated by neutron star radius measurements from low-mass X-ray binaries. For the supernova simulations which will be presented in the following, we have selected DD2.

The main characteristics of the HS EOS is that it combines a statistical ensemble of different nuclei within NSE with a RMF model for the unbound nucleons. The dissolution of nuclei at high densities is achieved by an excluded volume mechanism. A similar approach to construct supernova EOSs, based on a modified statistical model, has recently been published in Ref.~\cite{Furusawa:2011,furusawa2013}. Such statistical approaches are frequently used in the analysis of nuclear fragmentation reactions, where the so-called Statistical Multifragmentation Model is well established \cite{bondorf95}, and which also can be applied for supernova matter \cite{botvina10}. For a detailed comparison study of these three models, mainly focusing on heavy nuclei, see Ref.~\cite{Buyukcizmeci:2013}. 

Another important aspect of the subsaturation supernova EOS is the formation of light nuclear clusters, e.g., in the shock-heated matter. The generalized RMF (gRMF) model of refs.~\cite{Typel:2009sy,voskresenskaya2012} represents a very interesting new concept for the description of supernova matter with the emphasis on clusterization. This model utilizes the binding energy shifts due to Pauli-blocking obtained from the Quantum Statistical (QS) model (see e.g.\ Refs.~\cite{roepke82,roepke13}). The QS model itself can also be used to calculate the equation of state including light clusters. In the HS model, the quantum medium effects are mimicked by the excluded volume approach. In Ref.~\cite{hempel2011} it was shown that the abundances predicted with the HS model mostly lie inbetween the results of the gRMF and QS models, i.e. there is a good agreement for the prediction of the composition. However, thermodynamic quantities such as the free or internal energy show a moderate model dependency in the range 
where nuclear clusters are abundant. Another important approach for clusterized matter is given by the virial EOS \cite{horowitz06b}. At very low densities it is model independent, on the other hand, it cannot be applied at high densities, because it fails to describe the dissolution of nuclear clusters.

Another approach for the supernova EOS is presented in refs.~\cite{gshen2011a,gshen2011b}.  For low densities it uses the virial EOS. Here, the aforementioned problem does not occur, because intermediate densities are described with Hartree calculations of RMF interactions for a single representative heavy nucleus. For even higher densities close to normal nuclear matter density, uniform matter is obtained. Three different supernova EOS tables are already available for this model, namely for the RMF interactions FSUgold and NL3, and another one where FSUgold was phenomenologically modified, by adding artificially a pressure term in order to give a maximum neutron star mass of $2.1\:$M$_{\odot}$. This interaction is called FSU2.1. 

A detailed comparison between the HS and G.~Shen EOS has not been given so far in the literature. It would be very interesting, because it would allow to identify further the impact of the model description of nuclei, keeping the nucleon interactions unmodified. In Ref.~\cite{Steiner:2013} something similar was already done for the HS(TM1) and STOS EOS which are both based on TM1 interactions. However, such a comparison would be beyond the scope of the present article. Here, we concentrate on the effect in simulations of a new EOS which is consistent with most available constraints, namely HS(DD2), with the two standard models LS and STOS.

\subsection{Quark matter}

At densities on the order of several times $n_0$, the wave functions of individual nucleons start to overlap. As a consequence the description of nuclear matter composed of distinguishable nucleons could start to break down, resulting in a phase transition to the quark-gluon plasma. However, the conditions at which a phase transition may take place are currently highly uncertain. It can be constrained from heavy-ion experiments to some extend but the state of matter in heavy-ion collisions is intrinsically different to the one obtained in core-collapse supernovae. This is due to the large isospin asymmetry of matter and at least partial weak equilibrium. The search for a possible phase transition is part current and future heavy-ion experimental research at FAIR at the GSI/Darmstadt (Germany), NICA in Dubna (Russia), and RHIC in Brookhaven (US). 

In our study, we chose a representative quark matter EOS based on STOS for hadronic matter and the simple quark bag model for strange-quark matter. We select a bag constant of $B^{(1/4)} = 139\:$MeV and a strong interaction coupling constant $\alpha_s = 0.7$ (for more details see ref.~\cite{Sagert:2012} and references therein). The parameters are selected such that the resulting quark bag hybrid EOS (hereafter QB) has a hybrid star maximum mass of 2.04~M$_\odot$ and is thereby consistent with neutron star mass measurements~\cite{Demorest:2010,Antoniadis:2013}. The corresponding phase diagram features an extended quark-hadron co-existence region, i.e. a two phase mixture. The properties of the latter depend on the selected quark matter parameters and the chosen criterium for the construction of the mixed phase. In this approach,we apply the Gibbs construction. Under supernova conditions, i.e. temperatures on the order of tens of MeV and electron fractions of $Y_e\simeq0.2-0.3$, the critical density for the 
onset the two-phase mixture of hadronic and quark matter is close to saturation density.

\section{Characteristics and constraints of SN EOSs}

The saturation properties at $T=0$ for all mentioned hadronic supernova EOSs are listed in Table~\ref{tab:exp_prop1}, except for LS375, which is ruled out due to its too high value of the incompressibility. The EOS of STOS is based on the TM1 parameterization, and the EOSs of G.~Shen are based on NL3 and FSU. Thus we cover almost all nucleon interactions of existing supernova EOSs (for a detailed discussion of the saturation properties given in Table~\ref{tab:exp_prop1}, see ref.~\cite{Steiner:2013}). For additional theoretical and experimental constrains on the nuclear symmetry energy we refer to the other articles of this EPJA topical issue.

\begin{figure}[t!]
\includegraphics[width=1\columnwidth]{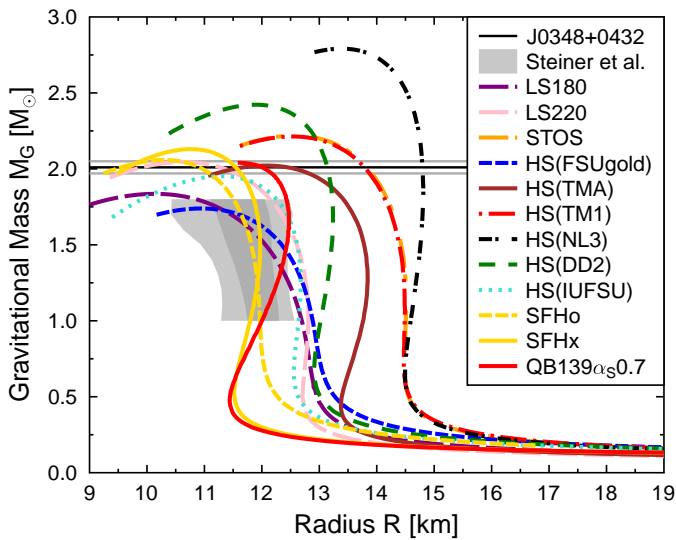}
\caption{Mass-radius relations for cold neutron stars in $\beta$-equilibrium for various different SN EOSs. (color online)}
\label{fig:mr}
\end{figure}

\begin{table*}[t]
\caption{Nuclear matter properties at saturation density, $n_0$, and zero temperature for our selection of hadronic SN EOS currently available. Listed are binding energy, $E_0$, incompressibility, $K$, symmetry energy, $S$, slope of the symmetry energy, $L$, radius of a 1.4~M$_\odot$ neuron star, $R_{1.4}$ and maximum gravitational mass, M$_\text{max}$.}
\begin{tabular}{c c c c c c c c}
\hline
\hline
& $n_0$ & $E_{0}$ &$ K$ &  $S$ & $L$ & 
 R$_{1.4}$ & M$_{\text{max}}$ 
\\
EOS & {[fm$^{-3}$]} & [MeV] & [MeV] &[MeV] & [MeV] & 
[km] & [M$_{\odot}$] \\
\hline
SFHo & 0.1583 & 16.19 & 245 & 31.57 & 47.10 & 11.89 & 2.06
\\
SFHx & 0.1602 & 16.16 & 238 & 28.67 & 23.18 & 11.99 & 2.13 
\\
HS(TM1) & 0.1455 & 16.31 & 281 &  36.95 & 110.99 & 14.47 & 2.21 
\\
HS(TMA) & 0.1472 & 16.03 & 318 &  30.66 & 90.14 & 13.85 & 2.02 
\\
HS(FSUgold) &  0.1482 & 16.27 & 229 &  32.56 & 60.43 & 12.55 & 1.74 
\\
HS(DD2) & 0.1491 & 16.02 & 243 & 31.67 & 55.04 & 13.22 & 2.42
\\
HS(IUFSU) & 0.1546 & 16.39 & 231 & 31.29 & 47.20 & 12.68 & 1.95 
\\
HS(NL3) & 0.1482 & 16.24 & 272 & 37.39 & 118.49 & 14.77 & 2.79 
\\
STOS(TM1) & 0.1452 & 16.26 & 281 &  36.89 & 110.79 & 14.50 & 2.22 
\\
\hline
LS (180) & 0.1550 & 16.00 & 180 & 28.61 & 73.82 & 12.16 & 1.84
\\
LS (220) & 0.1550 & 16.00 & 220 &  28.61 & 73.82 & 12.67 & 2.05
\\
\hline
Exp. & $\sim 0.15$ & $\sim 16$ & $240\pm10^1$ & $29.0-32.7^2$ & $40.5 - 61.9^2$ & $10.4-12.9^3$ & $\gtrsim 2.0^{4,5}$
\\
\hline
\hline
\end{tabular}
\\
$^1$ \cite{Piekarewicz:2010}
\\
$^2$ \cite{Lattimer:2013}
\\
$^3$ \cite{Steiner:2013}
\\
$^4$ \cite{Demorest:2010}, $1.97\pm0.04$~M$_\odot$
\\
$^5$ \cite{Antoniadis:2013}, $2.01\pm0.04$~M$_\odot$
\label{tab:exp_prop1}
\end{table*}
Note that the masses and especially the radii given in Table~\ref{tab:exp_prop1} are slightly different compared to the values of Ref.~\cite{Steiner:2013}, because we are using a different TOV-solver. Furthermore, there is a minor error in Ref.~\cite{Steiner:2013}: the radii of HS(TM1) and HS(TMA) were accidentally exchanged. Fig.~\ref{fig:mr} shows the corresponding mass-radius relations together with the constraints obtained from high-precision mass determinations~\cite{Antoniadis:2013}. Note that FSU and LS180 are not compatible with the mass limit of J0348+0432 \cite{Antoniadis:2013}. IUFSU was built to reach the mass limit of PSR J1614-2230~\cite{Demorest:2010} but is below the lower 1-$\sigma$ limit of the slightly more-massive NS which was reported recently in ref.~\cite{Antoniadis:2013}. All other SN EOS are compatible with the current maximum neutron-star mass constraint of $2.01\pm0.04$~M$_\odot$.

The determination of NS radii is still a very challenging task. Several groups obtain substantially different results\cite{Steiner:2010,Suleimanov11,Steiner:2013b,Guillot:2013,Lattimer:2013b} due to distinct model assumptions, e.g. composition and properties of the atmosphere. Nevertheless, a qualitative agreement of most studies points to moderate neutron-star radii for neutron stars with 1.4~M$_\odot$ which is also consistent with Chiral EFT~\cite{Hebeler:2010b,Steiner:2012}. For this reason, we include the results form the analysis of ref.~\cite{Steiner:2010} in Fig.~\ref{fig:mr} as a representative example. The simple non-linear RMF models TM1, TMA, and NL3, which do not contain additional meson couplings like FSUgold typically lead to large neutron star radii (see also Table~\ref{tab:exp_prop1}). The density-dependent RMF DD2 parameterization comes close (within 1~km) to the observational radius constraints. IUFSU was constructed to have both, small neutron-star radii like FSUgold and a large neutron 
star maximum mass (see Fig.~\ref{fig:mr}). The authors of SFHo and SFHx even extended this approach by fitting the EOS directly to the neutron star radius measurements. 
The two non-relativistic EOSs LS180 and LS220 are also compatible with small neutron star radii.

It is interesting to note that LS180 and FSU, as well as LS220 and IUFSU, have similar mass-radius curves. Nevertheless, the two LS models have very different  neutron matter EOSs, as will be shown below. Furthermore, the LS models lead to notable differences in core-collapse supernova simulations compared to FSUgold as was demonstrated in refs.~\cite{Steiner:2013,Hempel:2012}.

\begin{figure}[t!]
\includegraphics[width=1.\columnwidth]{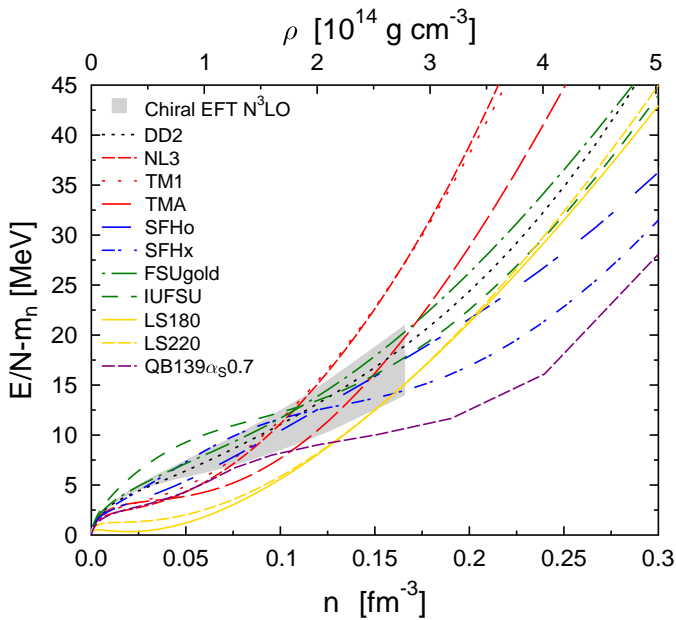}
\caption{Energy per baryon of neutron matter at zero temperature. The gray shaded region shows the results obtained with Chiral EFT from ref.~\cite{Krueger:2013}. The different lines show various available supernova EOS, for details see text. (color online)}
\label{fig:enm}
\end{figure}

Fig.~\ref{fig:enm} shows the energy per baryon, $E/N$, for neutron matter at $T=0$ for the same set of EOSs. The neutron matter EOS is important because its energy, $E/N$, gives a contribution to the nuclear symmetry energy, $S$. The slope of the curves is also important as it is directly related to the pressure $p$ via: 
\begin{equation}
p = n^2 \frac{\partial \left(E/N\right)}{\partial n}.
\end{equation}
Here, $n$ is the neutron number density. Note that the pressure of isospin symmetric nuclear matter is by definition zero at saturation density. Consequently, the pressure of neutron matter dominates the total baryon pressure around $\rho_0$.

Sophisticated new theoretical constraints for the neutron matter EOS became available in the last years. One of them is obtained from Chiral EFT. The latter represents a systematic approach to low density nuclear matter and allows to estimate theoretical error bars. The constraints from ref.~\cite{Krueger:2013} at N${^3}$LO are shown in Fig.~\ref{fig:enm} via the gray band. We remark that this band is consistent with many other up-to-date sophisticated models for neutron matter, for example Quantum Monte-Carlo~\cite{Gezerlis:2010}, Auxiliary Field Diffusion Monte-Carlo calculations~\cite{Gandolfi:2012}, or older variational calculations~\cite{Akmal:1998}.

The lines in Fig.~\ref{fig:enm} depict the different neutron matter EOSs. The used colors (color version online) distinguish the main characteristics of the underlying model for the bulk nucleon EOS. In yellow we present the results of the two LS non-relativistic Skyrme-like EOSs. They show significant deviations which were first noted in ref.~\cite{Krueger:2013}. The neutron matter EOS of LS180 is so soft that it even exhibits a region with negative neutron pressure where $d(E/N)/dn<0$. 

Red lines (NL3, TM1, and TMA) depict standard non-linear RMF models, where self-interactions of the $\omega$ and $\sigma$ mesons are included. These models experience problems in reproducing the results from Chiral EFT: At low densities they provide too much binding while at high densities they are too repulsive. The green lines (FSUgold, and IUFSU) show two RMF models where the $\omega-\rho$ coupling is included. Even though FSU was not fitted to neutron matter constraints it is in excellent qualitative agreement. However, its maximum neutron star mass is too small. For the construction of the IUFSU parameter set the authors modified FSU to obtain a sufficiently high neutron star maximum mass and fitted to the neutron skin thickness of $^{208}$Pb at the same time. As can be seen in Fig.~\ref{fig:enm}, IUFSU leads to the highest $E/N$ at densities below 0.1 fm$^{-3}$. One can conclude that the $\omega-\rho$ coupling is one possibility to obtain a reasonable behavior of the neutron matter EOS, even though it 
is difficult to obtain high enough maximum neutron star masses simultaneously.

The dotted black line shows DD2. It is the only relativistic SN EOS which is based on linear, but density-dependent couplings. The DD2 EOS has an excellent qualitative and quantitative agreement with Chiral EFT across all densities. Note that the parameterization DD, which is basically identical to DD2, (for details, see \cite{Typel:2005,Typel:2009sy}) has been introduced long before any of these constraints became available at the current precision. In the two models, SFHo and SFHx, shown by blue lines, various additional couplings and self-couplings are included. These two parameter sets have been determined only by charge radii and binding energies of finite nuclei and neutron star observations. Interestingly, they also give a better neutron-matter EOS than most of the other models. The polynomial ansatz of the couplings of ref.~\cite{steiner05} used in these two models is flexible enough to comply with various different EOS constraints, similar to the density-dependent approach. On the other hand, it has 
to be noted that the neutron-matter EOS of SFHx has some unexpected density dependence slightly below and up to saturation density (see Fig.~\ref{fig:enm}).

The purple dashed line in Fig.~\ref{fig:enm} shows the QB EOS, where the phase transition to strange quark matter sets in at about 0.07~fm$^{-3}$. Pure quark matter is reached at about $10 \times n_0$. The appearance of quark matter leads to pronounced differences to Chiral EFT for $E/N$. However, it is not clear if these constraints can be applied to an EOS with quark degrees of freedom. Note also that the quark densities within the two-phase mixture are generally higher than the total number density.

In conclusion, based on Chiral EFT, the neutron-matter EOS of LS and STOS as well as several other SN EOS can be classified as not compatible with recent constraints on the neutron matter EoS at low densities up to $n_0$, which also influences the density dependence of the symmetry energy. However, we remark that in SN matter trapped neutrinos prohibit extremely neutron-rich conditions with proton fractions $Y_p\ll 0.1$ and high temperatures. Therefore, it cannot be expected that a difference of a few MeV in the neutron-matter EOS has crucial consequences in core-collapse SN simulations. Nevertheless, the neutron matter EOS is an important aspect of nuclear matter and therefore these theoretical constraints should be taken into account. Note also that the presence of additional structures, for example nuclear clusters at sub-saturation densities and quark matter at super-saturation densities, is neither represented by the saturation quantities listed in Table~\ref{tab:exp_prop1} nor by the low-density 
neutron matter EOS in Fig.~\ref{fig:enm}. Such additional degrees of freedom may have a strong impact on neutron star data as well as on the supernova dynamics and observable signals, in particular when high temperatures and large isospin asymmetry are reached. 

\section{Results from core-collapse supernova simulations}

In the following paragraphs, we will discuss the impact of  the selected EOSs on the dynamics and the neutrino signal of core-collapse SNe. For this, we apply the $11.2\:$M$_\odot$ progenitor model from ref.~\cite{Woosley:2002zz}.

We will start our discussion with spherically symmetric simulations based on accurate neutrino transport. Below, we will briefly illustrate the differences to simulations in axial symmetry with spectral neutrino transport approximation.

\subsection{Simulations in spherical symmetry}

% Agile Boltztran brief %%%%%%%%%%%%%%%%%%%%%%%%%%%%%%%%%%%%%%%%%%%%%%%%%%%%%%%%%%%%%
Core-collapse supernova simulations in spherical symmetry are performed with the code AGILE-Boltztran. It is based on general relativistic radiation hydrodynamics and accurate three-flavor Boltzmann neutrino transport (see ref.~\cite{Liebendoerfer:2004} and references therein). For a list of implemented weak processes, see Table~I in ref.~\cite{Fischer:2012a}. For these spherically symmetric simulations, explosions could not be obtained for the simulated post bounce evolution up to 300~ms.

\begin{figure}[ht!]
\centering
\includegraphics[width=1.\columnwidth]{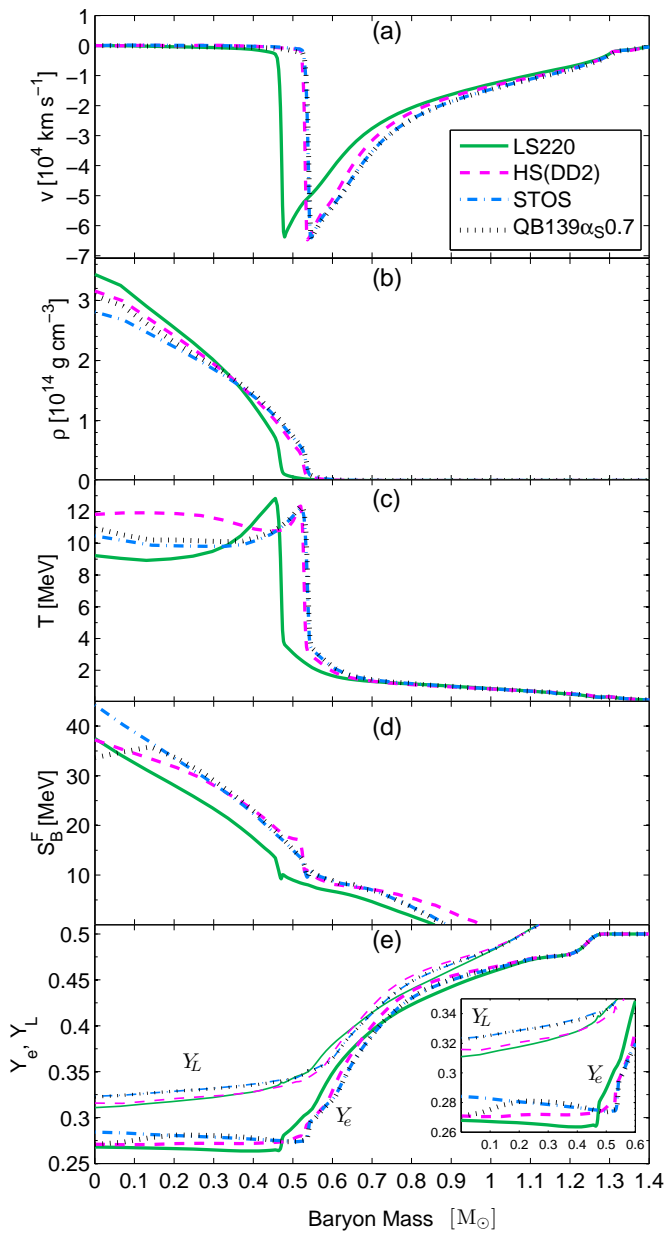}
\caption{(Color online) Bounce profiles with respect to the enclosed baryon mass of selected quantities, comparing the supernova EOSs LS220 (green lines), HS(DD2) (magenta lines), STOS (blue lines) and QB139$\alpha_S$0.7 (black lines).}
\label{fig:structure-bounce}
\end{figure}

Recently, improved rates for electron captures on heavy nuclei have become available~\cite{Juodagalvis:2010}. The authors calculate individual capture rates and spectra for several 1000 nuclear species. Finally, composition-averaged spectra and rates are provided by ref.~\cite{Juodagalvis:2010}, for which a distribution of the nuclear composition is assumed following free nuclear statistical equilibrium. It is valid for temperatures above $\sim0.5$~MeV and densities $\rho\leq 10^{13}$~g~cm$^{-3}$, above which ignoring nuclear interactions cannot be justified. The authors of ref.~\cite{Juodagalvis:2010} provide rates and spectra in tabular form in the three independent variables temperature, density and electron fraction, for which we apply linear interpolation. For details of the implementation, see ref.~\cite{Fischer:2013}. In the supernova simulations, the composition-averaged rates are then multiplied with the number density of heavy nuclei $n_{\langle A \rangle}$ obtained from the averaged 
nuclear composition given in terms of the average nuclear mass $\langle A \rangle$ provided by the HS EOS. Even though the ensemble considered in the rates of \cite{Juodagalvis:2010} may be slightly different than the one obtained with the HS EOS, it is at least obtained with a similar underlying description. Conversely, in the LS and STOS EOSs, only a representative heavy nucleus is considered, whose mass number can be systematically higher than in a statistical ensemble \cite{souza08}. Consequently, these EOSs are less consistent with the recent electron capture rates. In general, these rates represent an extension of the subset computed in ref.~\cite{Langanke:2003}. In comparison to the very simplified rates provided in ref.~\cite{Bruenn:1985en}, they result in generally lower central values of $Y_e$ and a different $Y_e$-profile towards lower densities at core bounce.

% core collapse phase %%%%%%%%%%%%%%%%%%%%%%%%%%%%%%%%%%%%%%%%%%%%%%%%%%%%%%%%%%%%%%%%%
The classical supernova EOSs, LS220 and STOS, have been widely discussed in the literature (see, e.g., \cite{Sumiyoshi:2006id}, \cite{Fischer:2009}), also in the context of multi-
dimensional supernova simulations~\cite{Marek:2008qi,Suwa:2013,Couch:2013}. In the preceding paragraphs, we compare simulations using the supernova EOSs LS220, STOS, HS(DD2) and the QB quark-matter EOS. Fig.~\ref{fig:structure-bounce} shows the bounce profiles as a function of the enclosed baryon mass of selected quantities, from top to bottom: velocity ($v$), density ($\rho$), temperature ($T$), free symmetry energy $(S_B^F)$, and  electron ($Y_e$) as well as lepton fraction ($Y_L$) in thick and thin lines respectively. Note that since in supernova simulations we are dealing with finite and even high temperatures, it is the \textit{free} symmetry energy which determines e.g. the state of $\beta$-equilibrium and not the internal symmetry energy. The free symmetry energy $S_B^F$ is defined via the expansion of the baryonic free energy per baryon in terms of the asymmetry, $\beta=1-2\,Y_e$, as follows
\begin{equation}
F_B(T,\rho,\beta) \simeq F_B^0 + \beta^2 S_B^F + \mathcal{O}(\beta^4) \; ,
\label{eq:fb}
\end{equation}
with $F_B^0=F_B(T,\rho,\beta=0)$. For uniform matter composed of only nucleons, and if the neutron-proton rest-mass difference is neglected, only even terms appear in the expansion above due to the exchange symmetry of neutrons and protons. Differentiating Eq.~(\ref{eq:fb}) with respect to $\beta$ and ignoring the higher order terms leads to:
\begin{equation}
\mu_n - \mu_p = 4\,\beta\,S_B^F(T,\rho)\,\,.
\label{eq:esym}
\end{equation}
Here, we use this expression to extract $S_B^F$ from the simulation results. The underlying expansion is a good approximation for homogeneous nuclear matter in the absence of nuclear clusters and low asymmetries. Contrary, the parabolic expansion of the EOS does not work well any more if their is a sizable contribution of nuclei, as was shown in Refs.~\cite{Typel:2009sy,typel2013}.

Electron captures on protons bound in nuclei determine the evolution during the stellar core contraction. They determine the conditions obtained at neutrino trapping which in turn sets the core lepton fraction $Y_L$. Since we use the same composition-averaged rates for electron captures in all simulations, the average nuclear composition, which is provided by the EOS, determines the core lepton fraction. I.e. fast deleptonization results in low $Y_L$, which is the case for LS220, while the opposite holds for STOS (see Fig.~\ref{fig:structure-bounce}). For LS220 the mass of the representative heavy nucleus $A$ is generally smaller than for STOS, and hence the electron capture rates are larger for LS220 than for STOS. Moreover, the nuclear composition of HS(DD2) differs only little from that of LS220, e.g., in terms of the average nuclear mass, and hence the core lepton fraction of HS(DD2) and LS220 differ only slightly. The Thomas-Fermi approximation of STOS leads generally to larger nuclei and hence, since 
the deleptonization proceeds slower than for LS220 and HS(DD2), it results in higher core $Y_L$.

\begin{figure}[ht!]
\centering
\includegraphics[width=1.\columnwidth]{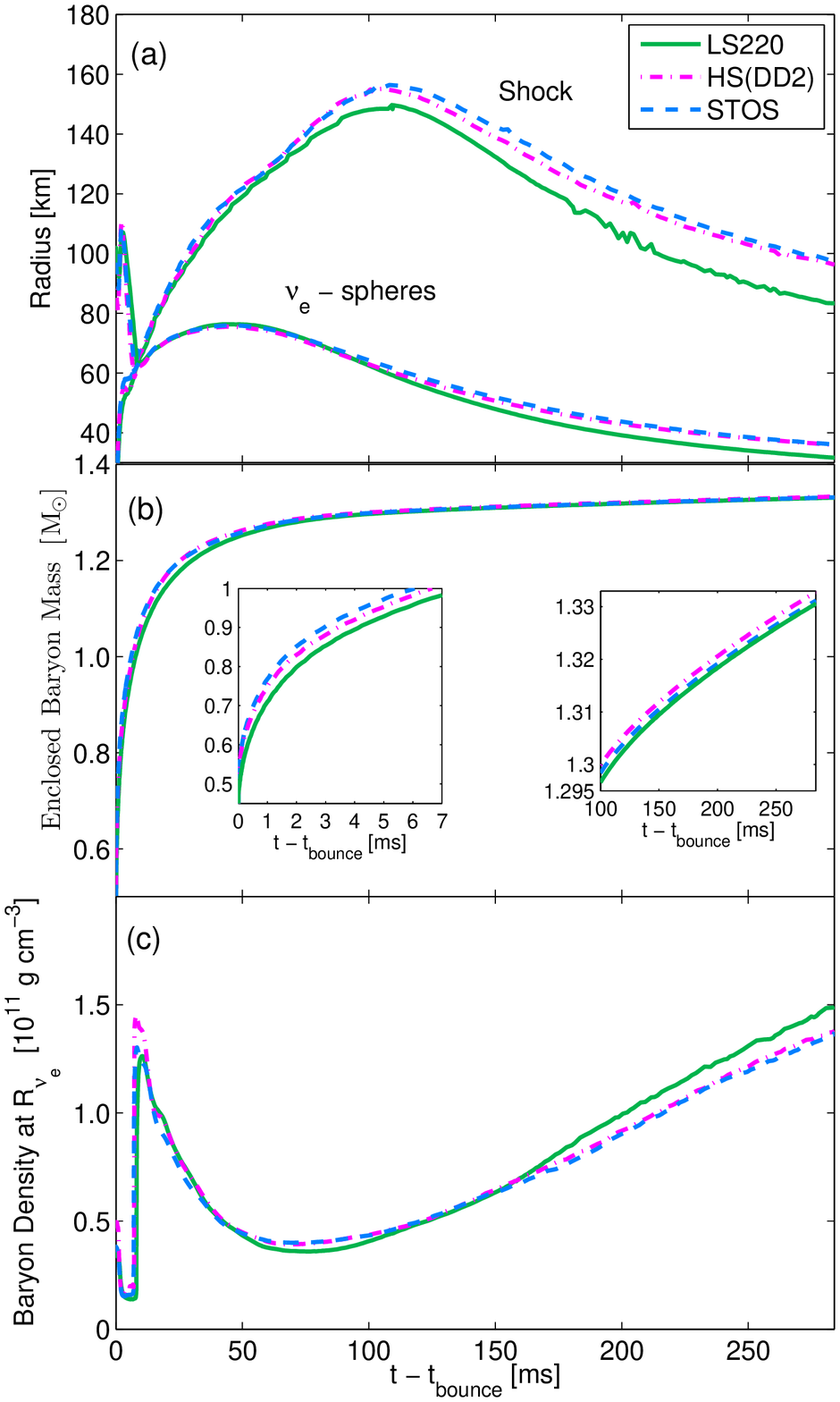}
\caption{Post-bounce evolution of selected quantities for the supernova EOSs LS220 (green), HS(DD2) (magenta), and STOS (blue). (color online)}
\label{fig:post-bounce_evolution}
\end{figure}

Beyond neutrino trapping, only a few milliseconds before core bounce, the further evolution of $Y_e$ is determined by the symmetry energy. Comparing STOS and LS220, the generally lower free symmetry energy of LS220 explains the overall lower electron fraction of the inner core. It results in higher central densities and less mass enclosed inside the shock at core bounce for LS220 (see Fig.~\ref{fig:structure-bounce}). At saturation density, HS(DD2) has a lower(higher) free symmetry energy than STOS(LS220). However, in supernova simulations also the density dependence of the symmetry energy $S_B^F(T,\rho$) is of relevance, and hence the slope of the symmetry energy. The less steep rise of $S_B^F$ at all densities (see Fig.~\ref{fig:structure-bounce}~(d)) is in agreement with a lower value of $L$ for HS(DD2), in comparison to LS220 and STOS (see Table~\ref{tab:exp_prop1}). At supersaturation densities, for HS(DD2) the rise of $S_B^F$ towards increasing densities reduces to a level where $S_B^F$ reaches values 
close to those of LS220. At subsaturation densities, the shallow slope of $S_B^F$ results in the largest symmetry energy for HS(DD2), slightly above that of STOS. Note that the different values of $S^F(T,\rho)$ could also be influenced by the different temperatures. Overall, the core symmetry energy for HS(DD2) lies between those of LS220 and STOS, and hence the core $Y_e$ for HS(DD2) has values between those of the two extreme EOS LS220 and STOS.  Consequently, the central density of HS(DD2) as well as the core density profile lay between those of LS220 and STOS. The different temperature profiles, also shown in Fig.~\ref{fig:structure-bounce}, are related to the different compactness and are partly given in terms of the different electron fraction profiles for LS220 and STOS. For a detailed discussion of the core-collapse phase comparing LS220 and STOS, see e.g. ref.~\cite{Hempel:2012}.

%\tf{To illustrate the differences in more details, Fig.~\ref{fig:structure-bounce}~(d) shows the  symmetry energy and the lepton fraction is illustrated in Fig.~\ref{fig:structure-bounce}~(e). It becomes clear that the different core $Y_L$ are related to the different evolutionary tracks during core contraction before complete neutrino trapping is reached, roughly above $\rho\simeq10^{13}$~g~ch$^{-3}$. It is determined from the deleptonization timescale, which in turn is given by the electron capture rates which incorporate the abundance of heavy nuclei. This is the main difference for the low- and intermediate-density EOS under investigation. Once neutrino trapping is achieved and $Y_L$ cannot drop any further, changes in $Y_e$ are now driven by the different symmetry energies of the EOSs under investigation and hence the core $Y_e$ scale with $S^F$.}

Recently, the possible appearance of quark matter in the supernova core received increasing attention (see, e.g.,  \cite{Takahara:1988yd,Gentile:1993ma,Sagert:2008ka,Fischer:2012b}). Therefore, we also show results obtained for simulations which include the QB EOS. This EOS was introduced above and is consistent with massive neutron stars. Up to the conditions for the appearance of quark matter, the core evolution during contraction and the core-bounce profile matches the one of the STOS simulation by construction (see Fig.~\ref{fig:structure-bounce}). Only above the critical density for the onset of strange quark matter differences occur. These are a slightly higher central density as a consequence of the lower core electron fraction obtained. The latter aspect is related to the lower symmetry energy of quark matter (see Fig.~\ref{fig:structure-bounce}). However, the core lepton fraction, which is determined at neutrino trapping at densities below the appearance of strange quark matter, is not affected. The 
symmetry energy of the QB EOS and its density dependence is also discussed in Fig.~3 of ref.~\cite{Fischer:2011}. These are important effects which are not covered by the saturation properties of nuclear matter at $T=0$. However, despite the extended quark-hadron mixed phase, only a slight softening of the high-density EOS is obtained in comparison to STOS. Moreover, the central region of the PNS where quark matter appears, remains stable up to several seconds after core-bounce. Pure quark matter is never reached. Initial expectations, that the PNS may undergo a second collapse resulting in the formation of a strong hydrodynamic shock wave as obtained in \cite{Gentile:1993ma,Sagert:2008ka,Fischer:2012b}, could not be fulfilled for the particular quark-matter properties of the QB EOS.

% post bounce evolution %%%%%%%%%%%%%%%%%%%%%%%%%%%%%%%%%%%%%%%%%%%%%%%%%%%%%%%%%%%%%%%%%%%%%
The post-bounce evolution during the first $300$~ms is shown in Fig.~\ref{fig:post-bounce_evolution}. Once the expanding bounce shock stalls, due to energy-losses from neutrino emission and the continuous dissociation of infalling heavy nuclei, the early post-bounce evolution is generally determined by mass accretion from the continuously gravitationally collapsing layers above the stellar core. It leads to a slow but continuous mass growth of the central PNS (given by the mass enclosed inside the shock), as illustrated in Fig.~\ref{fig:post-bounce_evolution}~(b). The sudden drop of the mass-accretion rate after about $100$~ms post bounce is due to the infall of the interface between Fe-core and Si-layer onto the shock, above which the baryon density is  significantly lower. The PNS growth-rate is generally determined by the softness of nuclear matter at both, high densities for the central properties of the PNS, and low/intermediate densities. The latter aspect is relevant for the compression of accumulated 
matter on the PNS surface. EOS differences obtained for the mass-growth rate are small, indicating a very similar compression behavior at low density during the considered timescales. The mass growth rates is mainly determined by the progenitor model. Larger differences are found for the evolution of the bounce shock and the neutrinospheres (see top panel in Fig.~\ref{fig:post-bounce_evolution}) after shock stalling at about $100$~ms post-bounce. These are mainly due to the large differences of the EOSs close to and above saturation density, which determines the central PNS contraction behavior (see Fig.~\ref{fig:enm}). The very soft LS220, with its extremely low symmetry energy (see Table~\ref{tab:exp_prop1}) at $n_0$ and lowest $E/N$ at sub-saturation densities (see Fig.~\ref{fig:enm}) leads to the fastest PNS contraction.

\begin{figure}[ht!]
\centering
\includegraphics[width=0.96\columnwidth]{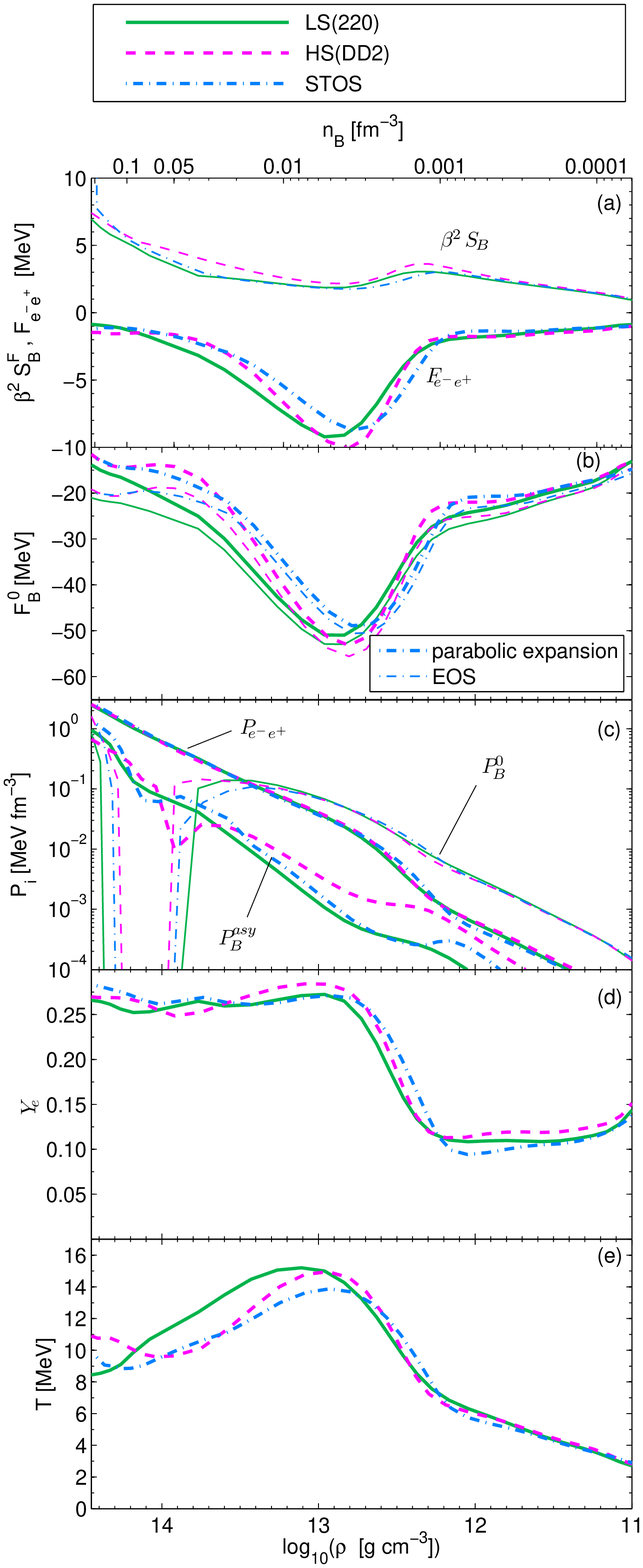}
\caption{(Color online) Selected quantities of the PNS, sampled at about 20~ms post bounce, comparing the supernova EOSs LS220 (green), HS(DD2) (magenta), and STOS (blue).}
\label{fig:esym_yl_pb_020}
\end{figure}
\begin{figure}[ht!]
\centering
\includegraphics[width=0.96\columnwidth]{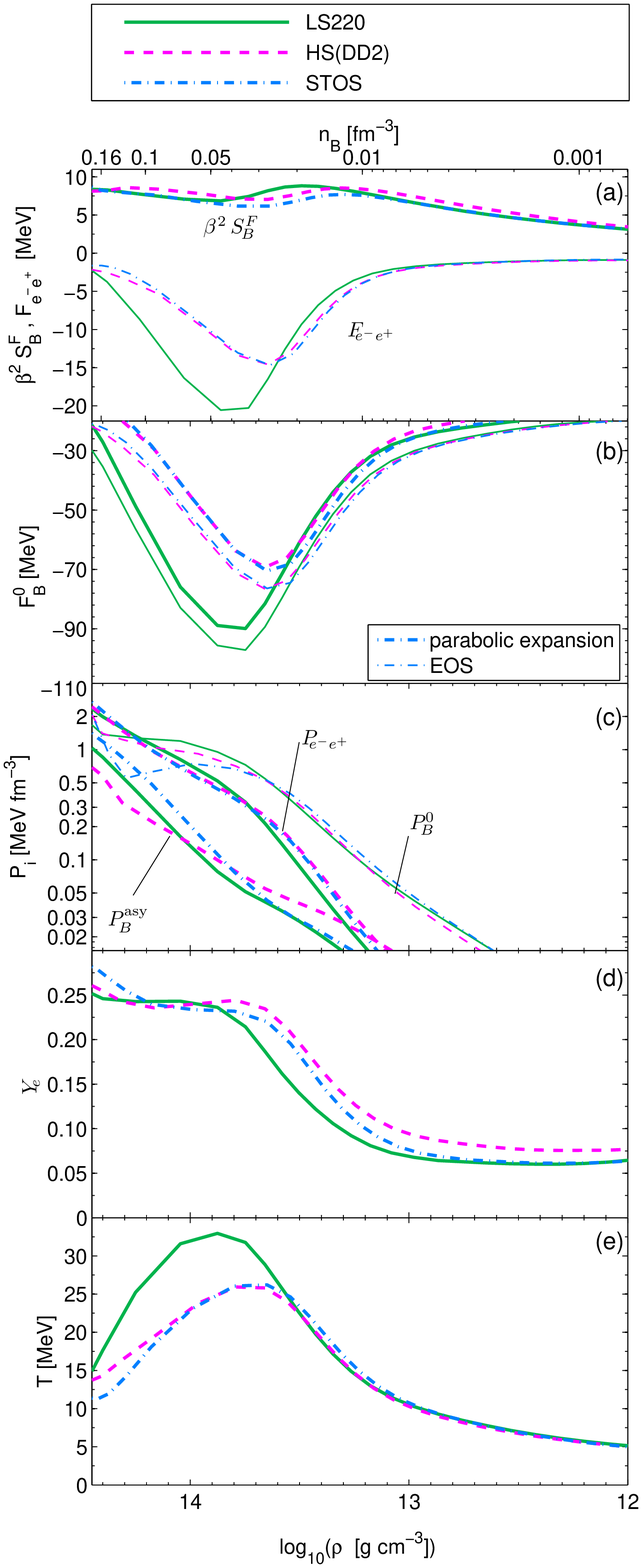}
\caption{(Color online) Selected quantities of the PNS, sampled at about 250~ms post bounce, comparing the supernova EOSs LS220 (green), HS(DD2) (magenta), and STOS (blue).}
\label{fig:esym_yl_pb}
\end{figure}

To this end, Figs.~\ref{fig:esym_yl_pb_020} and \ref{fig:esym_yl_pb} illustrate selected quantities relevant for the PNS evolution for the simulations with the different EOS LS220 (green), HS(DD2) (magenta), and STOS (blue) at about 20~ms and 250~ms post bounce respectively. The central $Y_e$'s (graphs (d) in Figs.~\ref{fig:esym_yl_pb_020} and \ref{fig:esym_yl_pb}) correlate with the symmetry energy , for which we show its contributions to the free symmetry energy, i.e. $\beta^2S_B^F$, in the graphs~(a) (thick lines) in the Figs.~\ref{fig:esym_yl_pb_020} and \ref{fig:esym_yl_pb}. Note that the total free energy, $F_\text{tot}(T,\rho,\beta) = F_B(T,\rho,\beta) + F_{e^-e^+}+F_\gamma$, has contributions from baryons, $F_B$, electrons/positrons, $F_{e^-e^+}$ and photons $F_\gamma$. We use the parabolic expansion Eq.~(\ref{eq:fb}) for the baryon free energy, decomposed into symmetric ($Y_e=0.5$), $F_B^0(T,\rho)$, and asymmetric parts ($Y_e\neq0.5$), $\beta^2\,S_B^F(T,\rho)$, both are shown in the graphs (a) and (
b) in the Figs.~\ref{fig:esym_yl_pb_020} and \ref{fig:esym_yl_pb}. Note that in the graphs (b) in the Figs.~\ref{fig:esym_yl_pb_020} and \ref{fig:esym_yl_pb} we also compare the symmetric part of the free energy, $F_B^0$, from the parabolic expansion (thick lines) with the full free energy obtained from the EOS by setting $Y_e\equiv 0.5$ (thin lines). The small differences are on the order of 5--10~MeV, which we attribute to the presence of nuclear clusters, whose effect in the EOS cannot be captured by the parabolic expansion of $F_B$. Moreover, the free energy defines the pressure, $P=n_B^2\partial F/\partial n_B|_{T,Y_e}$, which is the relevant quantity one has to compare when concluding about the evolution, e.g., the PNS contraction behavior as illustrated in Fig.~\ref{fig:post-bounce_evolution}. In the graphs (c) of the Figs.~\ref{fig:esym_yl_pb_020} and \ref{fig:esym_yl_pb} we show the individual contributions to the baryon pressure, again separated into symmetric $P^0$ and asymmetric parts $P^\text{
asy}$. Here we define $P^\text{asy}$ as the difference of the baryon pressure of asymmetric matter like in the simulation and of symmetric matter, i.e.$P^\text{asy}(T,\rho,Y_e)=P^B(T,\rho,Y_e)-P^B(T,\rho,Y_e=0.5)$.

In addition to electron captures, the symmetry energy determines mainly the evolution until core bounce, which holds also true for the early post bounce evolution up to about 5--10~ms. We find that until this moment the asymmetric baryon EOS, e.g. in terms of free energy and pressure, dominates over the symmetric EOS at all densities. Note that at even earlier times than shown in Fig.~\ref{fig:esym_yl_pb_020}, the asymmetric baryon EOS dominates over the symmetric, e.g., $P_B^\text{asy}\gg P_B^0$ at all densities. The situation changes only slowly as a consequence of the continuously rising temperature, starting in particular at low densities which is associated with the propagation of the bounce shock. As a consequence, the thermal contribution to the symmetric EOS becomes increasingly larger. At some point, the symmetric pressure $P_B^0$ rises not only above the asymmetric one $P_B^\text{asy}$ but it even dominates over the electron/positron gas $P_{e^-e^+}$, e.g., at low and intermediate densities as 
shown in Fig.~\ref{fig:esym_yl_pb_020} at about 20~ms after core bounce. Note that at high densities, the electron/positron contributions still  dominate over the baryon EOS. All EOS under investigation show qualitatively the same  behavior during the early post bounce phase.

The temperature increases continuously during the later post bounce evolution, as illustrated in Fig.~\ref{fig:esym_yl_pb}~(e). It is a direct consequence adiabatic compression induced from the continuous mass accretion prior to the shock revival, i.e. the explosion onset. Hence the importance of the symmetric contributions to the baryon EOS increases. At about 250~ms post bounce, the symmetric baryon EOS dominates over the asymmetric one at all densities (see graphs (a)-(c) in Fig.~\ref{fig:esym_yl_pb}). Only at very high densities, close to and above normal nuclear matter density, the values for symmetric and asymmetric baryon pressure contributions become of similar magnitude. The continuously decreasing electron/positron pressure, $P_{e^-e^+}$, which is also shown in Fig.~\ref{fig:esym_yl_pb}~(c), is related to the decreasing density and also to the decreasing $Y_e$ and $T$ with decreasing density. The symmetric part of the free energy as well as the electron/positron energy, are dominated by the 
temperature term (compare temperature curves in Fig.~\ref{fig:esym_yl_pb}~(e) and the $F$'s in Fig.~\ref{fig:esym_yl_pb}~(a)). Hence, the symmetry energy is only a perturbation during the later ($\sim 100$~ms) post bounce evolution. In this very aspect, supernova physics differs from that of cold neutron-stars, for which the symmetry energy dominates, e.g. the slope of the symmetry energy determines dominantly neutron star radii. Note that the central temperatures differ only by few MeV for all simulations (Fig.~\ref{fig:esym_yl_pb}~(e)). However, the fastest contraction is obtained for LS220, with hence highest peak temperatures and lowest central $Y_e$ (Fig.~\ref{fig:esym_yl_pb}~(d)). The less compact PNSs of the RMF EOSs HS(DD2) and STOS have significantly lower peak temperature, also shifted towards lower densities.

\begin{figure}[htp!]
\centering
\includegraphics[width=1.\columnwidth]{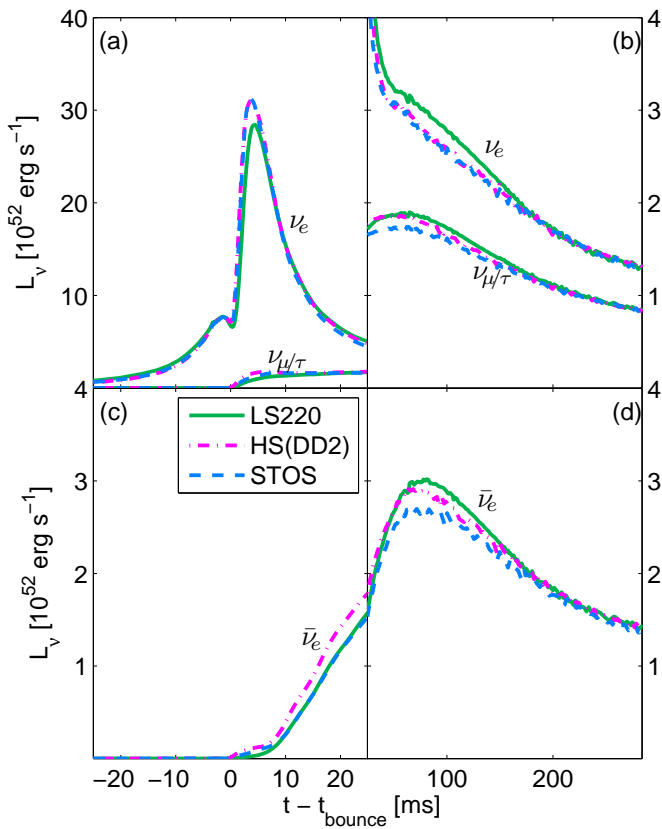}
\caption{Post-bounce evolution of the neutrino luminosities for the supernova EOSs LS220 (green), HS(DD2) (magenta), and STOS (blue). (color online)}
\label{fig:lumin}
\end{figure}
\begin{figure}[htp!]
\centering
\includegraphics[width=1.\columnwidth]{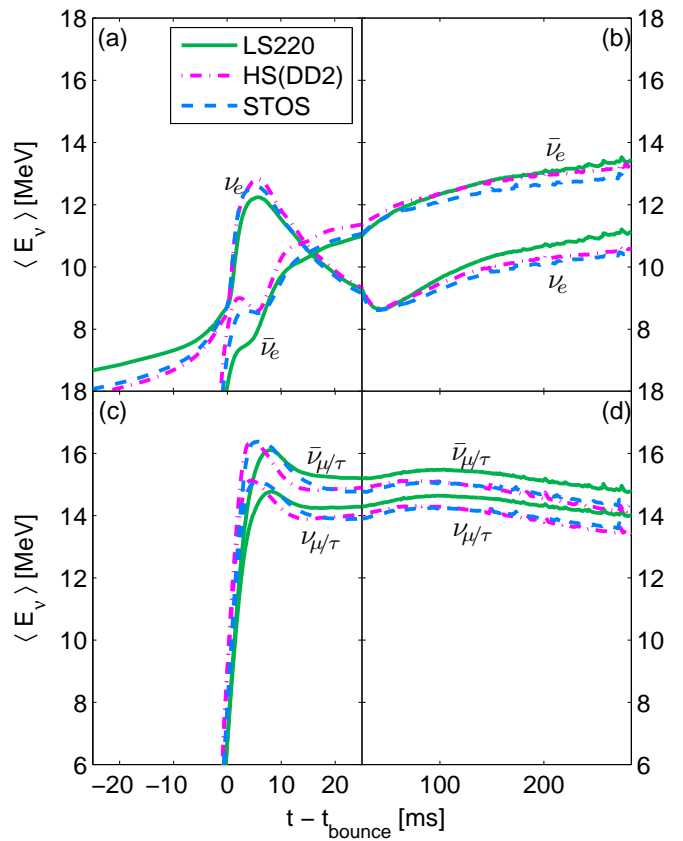}
\caption{Post-bounce evolution of the average neutrino energies for the supernova EOSs LS220 (green), HS(DD2) (magenta), and STOS (blue). (color online)}
\label{fig:energy}
\end{figure}

Note that the properties around saturation density are of relevance for the central compression behavior of the PNS, even if the densities exceed $n_0$. The simulation using the very stiff STOS, with a very large symmetry energy and a high value of $E/N$, results in the slowest PNS contraction. Our choice of the optimal EOS HS(DD2) lies in between LS220 and STOS. Note that neither LS220 nor STOS are in the overall acceptance range for the neutron-matter EOS predicted from Chiral EFT. With that, HS(DD2) has an optimal density dependence (see Fig.~\ref{fig:enm}). This shows that not only nuclear matter properties, such as the symmetry energy, at saturation density are of importance but also their density dependence. Moreover, conclusions which are drawn on supernova dynamics from the saturation properties of nuclear matter at $T=0$ apply only partially because supernova matter, in particular inside the PNS, is isospin asymmetric and has finite and even high temperatures. At low densities, neutrino decoupling 
and hence neutrino cooling/heating takes place (see the evolution of the neutrinospheres in the top panel of Fig.~\ref{fig:post-bounce_evolution} and the density at the neutrinospheres in the bottom panel). Differences between HS(DD2) and STOS are small and can be related to the different nuclear matter properties at very low densities, originating from a different description of nuclei. Only LS220, with the rapid PNS contraction, leads to significantly higher densities and also a much more compact PNS with a higher central density and peak temperature. This may be due to the extremely low symmetry energy at $\rho_0$ as well as neutron-matter energy $E/N$ at sub-saturation densities (see Fig.~\ref{fig:enm}).

% new: neutrino discussion
The corresponding evolution of neutrino luminosities and average energies for the hadronic EOSs LS220, HS(DD2), and STOS, are illustrated in Figs.~\ref{fig:lumin} and \ref{fig:energy}, respectively. The quantities are sampled in the co-moving frame of reference at a radius of 1000~km. Detailed comparisons, in particular between LS and STOS, have already been provided by~\cite{Sumiyoshi:2007pp,Fischer:2009,OConnor:2011,Hempel:2012,Steiner:2013}. Differences obtained during the core collapse phase, as well as for the deleptonization burst (see Figs.~\ref{fig:lumin} (a), (b) and \ref{fig:energy} (a), (b)), are related to the different nuclear composition. However, in view of the possible shock revival after shock stalling on timescales on the order of several $100\:$ms, differences obtained during the post-bounce evolution play a more important role. The average energies obey a clear hierarchy with $\langle E_{\nu_e} \rangle < \langle E_{\bar\nu_e} \rangle < \langle E_{\nu_{\mu/\tau}} \rangle$. This reflects 
the different neutrino decoupling regions ($\nu_e$: lowest density, $\bar\nu_e$: higher density due to different $Q$-value for charged-current reaction, $\nu_{\mu/\tau}$: highest density) resulting from weak processes that contribute to the corresponding neutrino flavors (for details, see ref.~\cite{Raffelt:2001,Keil:2003,Fischer:2012a} and references therein). The fast(slow) PNS contractions, due to the soft(stiff) EOS LS220(STOS) result in high(low) average neutrino energies (see Figs.~\ref{fig:energy} (b), (d)). The faster PNS contraction for LS220 is also reflected in the steeper slope of the luminosity (see \ref{fig:lumin} (b), (d))), indicating faster retracting neutrinospheres at the PNS surface. This is related to the fastest drop in the mass accretion rate for supernova simulations using LS220 (in comparison to those with STOS and HS(DD2)). This effect is most pronounced for the electron (anti)neutrinos which decouple at lowest densities. The PNS contraction of the novel HS(DD2) EOS lies in between 
the LS220 and STOS EOSs regarding the evolution of the average energy and luminosity for the entire post-bounce phase.

\subsection{Supernova explosions in axial symmetry}

The axially symmetric supernova simulations discussed here are based on Newtonian radiation hydrodynamics. It employs the ZEUS-2D hydrodynamics code~\cite{Stone:1992} and neutrino radiative transfer for $\nu_e$ and $\bar\nu_e$ using the Isotropic Diffusion Source Approximation (IDSA). It is well calibrated to reproduce the results of full Boltzmann transport during the accretion phase prior to the possible onset of an explosion~\cite{Liebendoerfer:2009a}. For details about the supernova model, see refs.~\cite{Suwa:2009py,Suwa:2011,Suwa:2013}. In addition to axially symmetric simulation, we have performed fully three-dimensional neutrino-radiation-hydrodynamic simulations~\cite{Takiwaki:2012}. Here, we compare the two EOS LS220 and STOS, for both of which neutrino-driven explosions were obtained aided by convection and the standing accretion shock instability (SASI). Note that differences between Newtonian and fully relativistic simulations have been discussed in detail in ref.~\cite{Liebendoerfer:2001b} 
based on the spherically symmetric case. 

\begin{figure}[ht!]
\centering
\includegraphics[width=1.\columnwidth]{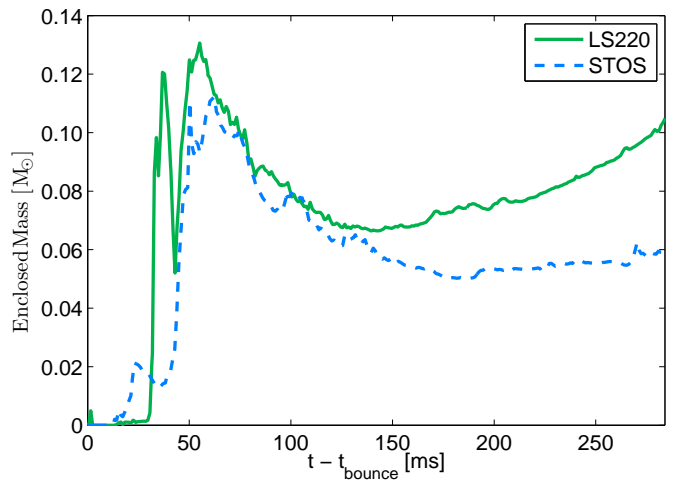}
\caption{Post-bounce evolution of the enclosed mass inside the gain region for the supernova EOSs LS220 (green solid line) and STOS (blue dash-dotted line). (color online)}
\label{fig:gain_2d}
\end{figure}

In multi-dimensional supernova simulations, convection and hydrodynamic instabilities, which are driven by neutrino heating and cooling, dominate the post bounce evolution. Consequently, differences to the spherical case can be very large and the above reported differences due to the nuclear EOS may be altered~\cite{Marek:2008qi,Suwa:2013,Couch:2013}. Currently available multi-dimensional supernova simulations indicate a structural feedback of the PNS to the SN dynamics at lower densities. It relates to the mass enclosed inside the gain region, for which the evolution is shown in Fig.~\ref{fig:gain_2d} comparing LS220 and STOS for the same 11.2~M$_\odot$ progenitor as discussed above. The softer LS220 leads to significantly more mass enclosed inside the heating region than the stiffer STOS (for details, see ref.~\cite{Suwa:2013}). This, in turn, leads to larger heating and a more optimistic situation for the onset of a neutrino driven explosion for LS220, which is contrary to the spherically symmetric 
simulations.

\begin{figure}[ht!]
\centering
\includegraphics[width=1.\columnwidth]{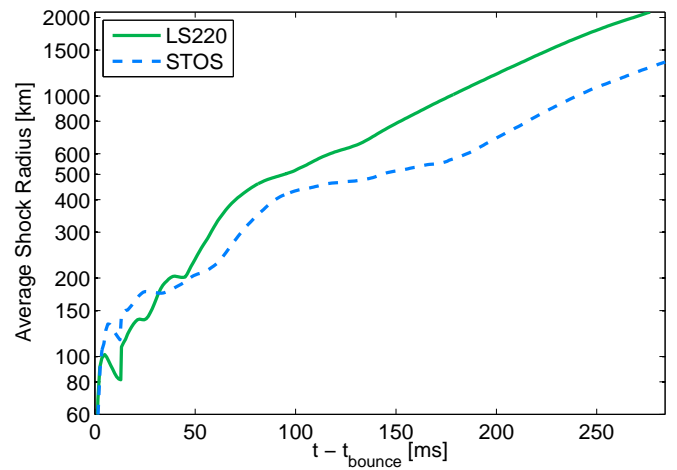}
\caption{Post-bounce evolution of the average shock position for the supernova EOSs LS220 (green solid line) and STOS (blue dash-dotted line). (color online)}
\label{fig:shock_2d}
\end{figure}

The difference of these simulations comes from how the PNS contracts. The faster contraction leads to the stronger pressure wave generation around the surface of PNS. The pressure wave hits the shock wave and transfers the momentum to the shock, such that the shock wave propagates outwards. This feature is well demonstrated in ref.~\cite{Suwa:2013} using a different progenitor model. As mentioned in the previous subsection, LS220 implies the faster contraction of the PNS than that of STOS, thus our axially symmetric simulation of LS220 actually indicates better condition for explosion (see also ref.~\cite{Janka:2012}).

Up to about 50~ms post bounce, the axially symmetric simulations agree qualitatively with the spherically symmetric case, as illustrated via the shock evolution in Fig.~\ref{fig:shock_2d}.  The simulations using STOS seem generally more optimistic for the possible onset of an explosion than for LS220, i.e. a larger shock radius.  However, the larger mass inside the heating region for LS220 leads to an earlier onset of the shock expansion than for STOS. This is aided by neutrino-driven convection and the development of SASI. This is the case $>50$~ms post bounce (see the shock evolution in Fig.~\ref{fig:shock_2d}). Even after the explosion onset, the larger heating for LS220 remains and leads to a faster shock expansion to increasingly larger radii than for STOS.

Note that our axially symmetric simulations omit heavy flavor neutrinos and the related energy loss. This, in combination with Newtonian gravity, may be responsible for the very early onset of explosion in comparison to simulations that include more sophisticated microphysics~\cite{Mueller:2012,Bruenn:2013}.
\begin{figure*}[ht!]
\centering
\subfigure[\,\,Mass accretion phase.]{
\includegraphics[width=0.96\columnwidth]{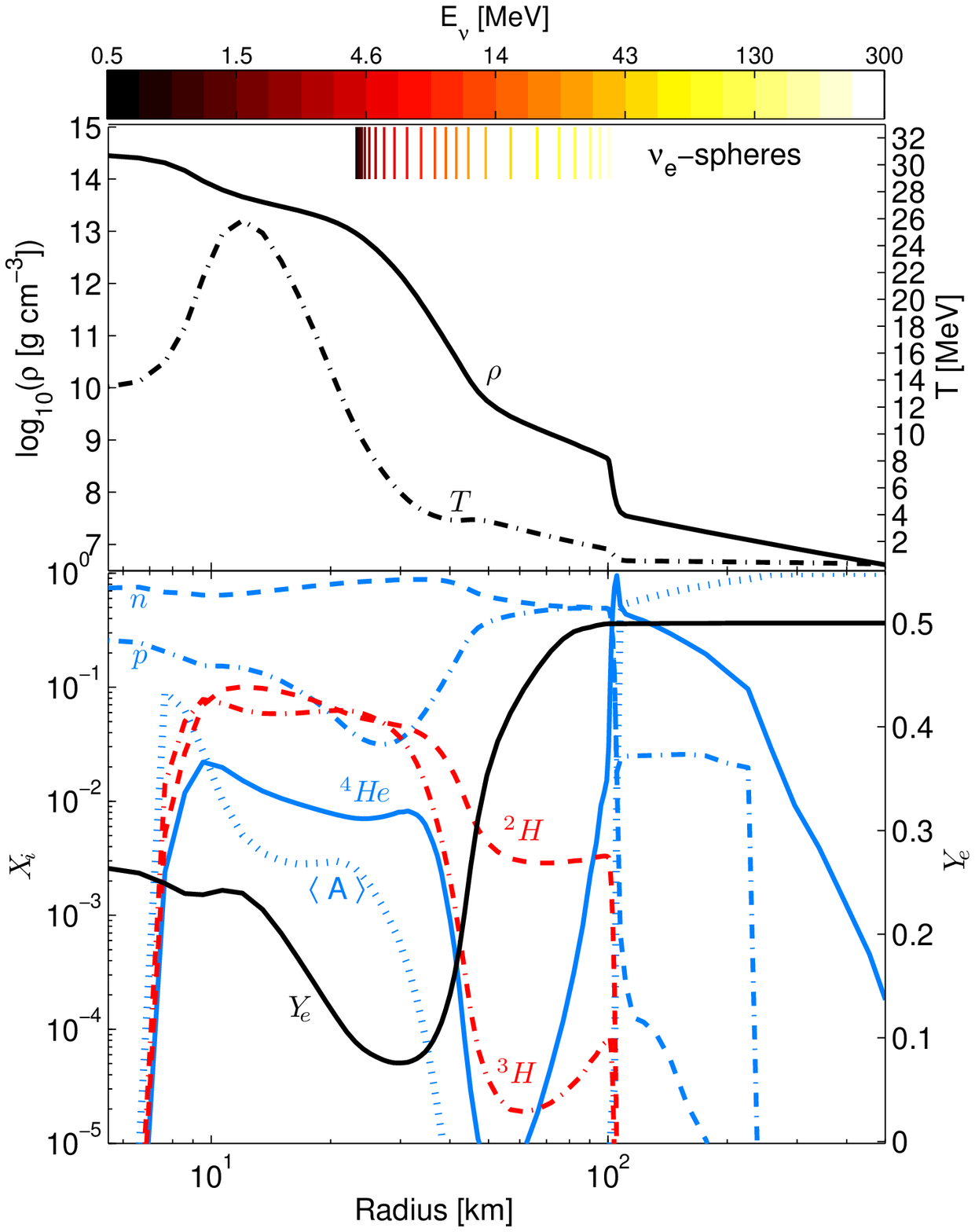}
\label{fig:cluster_sn_a}}
\hfill
\subfigure[\,\,PNS deleptonization phase.]{
\includegraphics[width=0.96\columnwidth]{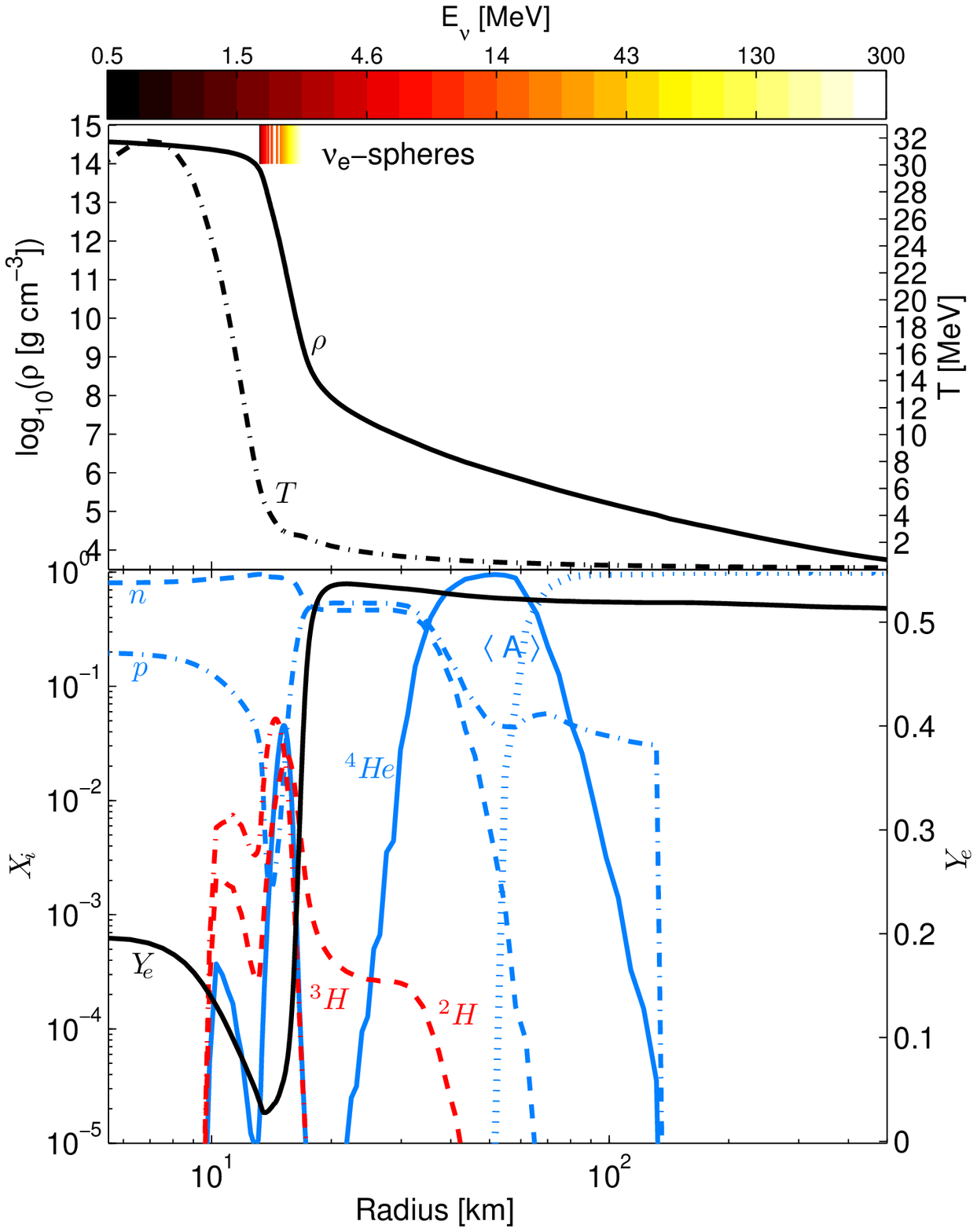}
\label{fig:cluster_sn_b}}
\caption{Mass fractions of free nucleons (blue), $\alpha$-particles (blue), heavy nuclei $\langle A \rangle$ (blue) and other light clusters (red) as well as $Y_e$ (black) at two selected conditions relevant for core-collapse supernova studies, prior to a possible explosion onset at 250~ms post bounce (a) and at about 5~seconds after the explosion onset (b). The top panels show the corresponding radial profiles of baryon density and temperature. The region of neutrino decoupling is illustrated via the energy-dependent neutrinospheres where color-coding indicates the neutrino energy. (Color online)}
\label{fig:cluster_sn}
\end{figure*}

\subsection{Light nuclear clusters in supernova simulations}

One aspect which has received increasing attention during the last years is the presence of light nuclear clusters and their potential impact (see, e.g., refs.~\cite{Sumiyoshi:2008,Arcones:2008,Typel:2009sy,hempel2011,furusawa2013}). The situation is illustrated in Fig.~\ref{fig:cluster_sn}, where we plot radial profiles of the mass fractions of the classical supernova composition, free nucleons and helium ($^4$He), as well as the light nuclear clusters deuteron ($^2$H) and triton ($^3$H). The latter two are only included in HS(DD2). The results that will be discussed in the following are based on HS(DD2) which takes into account on the order of several 1000 heavy nuclear clusters as well as all light nuclear clusters. The spherically symmetric supernova simulations are performed with AGILE-Boltztran and use the same $11.2\:$M$_\odot$ progenitor star as above.

Fig.~\ref{fig:cluster_sn_a} represents typical conditions for the post-bounce mass accretion phase, during which the gradients of density and temperature at the PNS surface are relatively shallow (see the top panel). In addition, the neutrino spectra and luminosities are determined from charged-current processes which take place in the continuously accumulated material on the PNS surface. Hence, depending on the neutrino energy, the neutrino decoupling region spans over a large distance up to the standing accretion shock at about 100~km (see top panel). Light nuclear clusters exist only in the high-entropy dissociated regime behind the standing bounce shock and their abundance is low at small densities. With increasing baryon density, their amount increases reaching the level of free protons (see bottom panel in Fig.~\ref{fig:cluster_sn_a}). At and above saturation density, the abundances of  clusters decrease again and homogeneous matter is reached. Note, that the nuclear composition close to the energy-
averaged neutrinospheres (see top panel in Fig.~\ref{fig:cluster_sn_a}) is dominated by free nucleons. The average neutrino energies are between $10-15\:$MeV. Deuterons and tritons may affect only the low-energy neutrinos with energies of $0.5-5\:$MeV as these decouple at highest densities where $^2$H and $^3$H are as abundant as protons~(see Fig.~\ref{fig:cluster_sn_a}). However, these neutrinos have a negligible impact on the total energy-loss that dominates at these densities of the cooling region. Moreover, the weak processes with deuterons and tritons which determine the energy loss are highly suppressed due to the large $Q$-value. In the heating region behind the bounce shock, densities are significantly smaller with mass fractions of deuterons and tritons being lower by several orders of magnitude. Hence, a strong impact of light nuclear clusters on neutrino heating/cooling and thereby on the supernova dynamics cannot be expected prior to the possible explosion onset.

The situation changes once the standing accretion shock has been revived. The latter determines the onset of the supernova explosion. To model this phase for an evolution up to several seconds after the explosion onset, we apply the spherically symmetric SN code AGILE-Boltztran and enhance the neutrino heating/cooling rates in order to trigger the explosion. Once the shock has been revived, we switch back to the standard rates (for details, see~\cite{Fischer:2009}). The mass fractions of free nucleons, alpha particles, and other light clusters are shown in the bottom panel of Fig.~\ref{fig:cluster_sn_b}, at about 5~s after the explosion onset. At this point in the simulation  mass accretion vanishes and the PNS settles into a quasi-stationary state. As a consequence, the gradients of density and temperature at the PNS surface steepen significantly (see top panel). Moreover, the neutrino spectra are no longer determined by mass accretion but are increasingly dominated by neutral-current processes and 
therefore represent diffusion spectra. Their neutrino decoupling shifts to significantly higher densities and spreads over a large range of densities, with a very small radial range. In this region, matter is very neutron rich with $Y_e=0.05-0.3$. The conditions favor the presence of light nuclear clusters making them more abundant than free protons by one order of magnitude (see bottom panel of Fig.~\ref{fig:cluster_sn_b}). On long timescales of $10\:$s this may influence the deleptonization of the PNS via weak-processes with the abundant light clusters $^2$H and $^3$H. The importance of clusters in supernova simulations has been discussed in refs.~\cite{Sumiyoshi:2008,Arcones:2008,furusawa2013}. Clusters may also leave an imprint in the neutrino signal and the consequent nucleosynthesis of heavy elements in the neutrino-driven wind which is ejected form the PNS surface via continuous neutrino heating after a successful explosion. The description of this phase requires a consistent implementation of weak 
processes and the nuclear EOS, i.e. taking into account medium modifications for charged and neutral current weak rates with nucleons~\cite{Fischer:2012a,MartinezPinedo:2012,Roberts:2012}. These medium modifications of the vacuum $Q$-value are related to the nuclear symmetry energy. Note that when implementing such weak processes with e.g. $^2$H and $^3$H in supernova codes, it is important to consider not only final-state Pauli blocking for both nucleons and electrons/positrons but also the medium modifications of the vacuum $Q$-values. The latter will dominate the energetics of the weak processes with light clusters at high densities ($\sim10^{13}-10^{14}$~g~cm$^{-3}$) where these are as abundant as protons. Generally, this leads to a suppression of the low-energy neutrinos.

\section{Symmetry energy impact on cluster formation in supernovae}

\begin{figure}[ht!]
\centering
\includegraphics[width=1.\columnwidth]{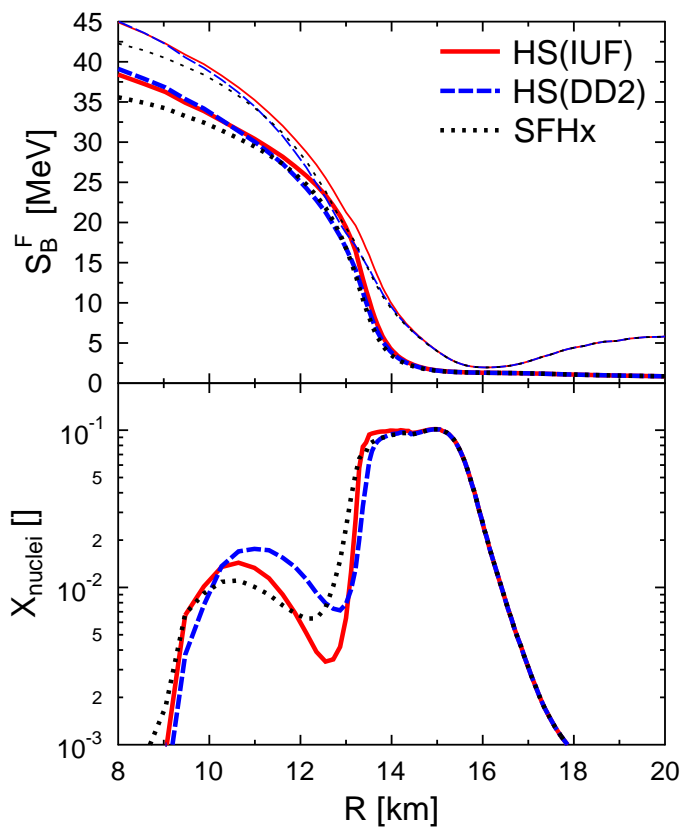}
\caption{Free symmetry energy (top panel) and mass fraction of nuclei (bottom panel) comparing the different EOS HS(IUF) (red solid lines), HS(DD2) (blue dashed lines), and 
SFHx (black dotted lines). We compare two different definitions $S_B^F$, the difference of the free energy per baryon of neutron matter and symmetric matter (thin lines) compared with the free symmetry energy of only nucleons (thick lines). The profiles are those from the supernova explosion simulation at about 5 seconds after explosion onset shown in Fig.~\ref{fig:cluster_sn_b}. (color online)}
\label{fig:sym_lights}
\end{figure}

As mentioned above, the formation of nuclei gives a contribution to the symmetry energy. However, it is also interesting to ask for the effect of the symmetry energy of uniform nuclear matter, i.e., nucleons, on the abundances of nuclei. To examine this effect we show the summed mass fractions of nuclei $X_\text{nuclei}=1-X_n-X_p$ in Fig.~\ref{fig:sym_lights}, evaluated with three different EOSs: HS(IUF), HS(DD2) and SFHx. We step back from comparison of the standard supernova EOS LS220 and STOS here, because within their simplified description of the nuclear composition light nuclear clusters are not taken explicitly into account. The conditions, i.e. density, temperature and $Y_e$ profiles, are taken from the simulation at 5~seconds after explosion onset as shown in Fig.~\ref{fig:cluster_sn_b}. The upper panel in Fig.~\ref{fig:sym_lights} depicts two different definitions of the free symmetry energy. The thin lines are the difference of the free energy per baryon of neutron matter and symmetric matter at 
the temperature and density given by the simulation. The thick lines are the free symmetry energy of only nucleons, defined by the second derivative of the free energy with respect to asymmetry of the baryon EOS. It is evaluated for the partial density of the nucleons as given by the full EOS.

By comparing the thick and thin lines one can identify the contribution of non-quadratic terms in the asymmetry expansion of the EOS, and most importantly also the effect of nuclei. For radii above 18~km, the mass fraction of nuclei drops below $10^{-3}$. Nevertheless, the two different definitions of the symmetry energy do not agree, because for $Y_e=0.5$ there would be a sizable contribution of nuclei. Also for very high densities (i.e.\ low radii) the two definitions of the symmetry energy give different results, mainly because of the quartic contribution of the kinetic part of the free symmetry energy of nucleons.

Moreover, the different nucleon interactions have also an impact on the abundance of nuclei. For radii larger than 14~km, basically all EOSs predict the same composition, because the densities are so low, $n_B \lesssim 0.005$~fm$^{-3}$, that the nucleon interactions are almost negligible. Furthermore, above 9~km the densities are in excess of saturation density and temperatures are so high that nuclei are dissolved. In the intermediate range between 9 -- 14 km, the abundances from the different EOS can differ by several percent, however the overall qualitative behavior is not changed. Interestingly, the differences found cannot be related to the symmetry energy. Consider e.g. the local minima in $X_\text{nuclei}$ of HS(IUF) around 12.5~km. It occurs in a region, where HS(IUF) provides the highest symmetry energy, which is connected to its high energy of the neutron matter EOS at subsaturation density pointed out before. Intuitively one would expect that a high symmetry energy of nucleons leads to an 
enhancement of the abundance of nuclei, but here the opposite is the case. This means in turn that the different mass fractions are more related to the differences in the isoscalar part of the nucleon interactions, which is in qualitative agreement with Ref.~\cite{ferreira2012}.

A reason for the little effect of the symmetry energy is simply given by the fact, that the symmetry energies of the three models in the range of 9 to 14~km are rather similar. Only for higher densities differences become significant, where nuclei are not abundant any more. Furthermore, it is well known that fitting of nuclear energy density functionals to binding energies of nuclei leads to a ``fix-point'' of the symmetry energy around 2/3 -- 3/4 of saturation density, i.e. where all such models give similar predictions. In Fig.~\ref{fig:sym_lights} this density is reached around 11.5~km. We can confirm that the free symmetry energies at this radius are indeed all very similar. 

We conclude that (within the HS model) the nucleon interactions, and in particular the symmetry energy of nucleons, have only a minor effect on the abundances of light nuclei in the PNS star envelope. In Ref.~\cite{hempel2011} one of the authors compared already the predictions for the cluster abundances of the HS EOS with quantum statistical model and found overall good agreement. A comparison with the virial EOS would be interesting. However note that here one is close to the high-density regime where this EOS cannot be applied any more, because it cannot achieve the dissociation of light clusters as observed in Fig.~\ref{fig:sym_lights} below 9~km for the HS EOS.

Note that in the analysis we did not take into account the effect of the EOS in the simulation, e.g. that different EOSs will lead to different asymmetries, as discussed in the previous sections.

%%%%%%%%%%%%%%%%%%%%%%%%%%%%%%%%%%%%%%%%%%%
\section{Summary}

In this article we have reviewed a comprehensive selection of currently used supernova EOSs, all of which differ in their nuclear matter properties. From Chiral EFT, which provides currently a sophisticated description of nuclear matter up to saturation density at zero temperature, most of their associated neutron-matter EOSs can be ruled out. It includes in particular the classical and most widely used EOSs LS and STOS, although they are partly consistent with current constraints of low-mass neutron star radii and maximum neutron star masses. The EOS which currently satisfies most of the nuclear as well as astrophysical constraints are DD2 and SFHo~\cite{Steiner:2013}, whereas the latter was not used in the present study (see also ref.~\cite{Lattimer:2013}). Note that even in cases of very similar mass-radius relations for different EOS, their nuclear matter properties including the neutron matter energy per baryon can be very different. Within the comparison of IUFSU and LS220 we found that extremely different 
neutron matter EOS at sub-saturation densities have very similar mass-radius relations. It shows that for astrophysics, the most relevant EOS differences occur around saturation density and above, which is consistent with the well known importance of the slope of the symmetry energy $L$ for neutron star matter.

We apply a selection of EOSs in core-collapse supernova simulations of a massive Fe-core progenitor of 11.2~M$_\odot$ in spherical symmetry. These are based on general relativistic radiation hydrodynamics and three-flavor Boltzmann neutrino transport. We examine the obtained differences, such as the conditions at core bounce, and illustrate the early post-bounce evolution prior to the possible onset of an explosion, comparing LS220, STOS, and HS(DD2).

During core collapse, the deleptonization is determined by the electron capture rate~\cite{Bruenn:1985en,Langanke:2003ii,Hix:2003}. The rates which involve heavy nuclei scale with the inverse average nuclear mass $\langle  A \rangle$ which is provided by the EOS. Hence, we find the fastest(slowest) deleptonization for EOS with smallest(largest) $\langle  A \rangle$ such as LS220(STOS) and hence a low(high) $Y_L$ at neutrino trapping. Once neutrino trapping is reached and $Y_L$ changes no more, the further evolution of $Y_e$ is determined by the symmetry energy. I.e., a large(low) symmetry energy results in a high(low) $Y_e$. Since HS(DD2) and LS220 have very similar nuclear composition, core $Y_L$ and $Y_e$ are more similar than those of STOS. Note that not only the nuclear matter properties at saturation density are of relevance but also their density dependence at finite temperatures.

The post bounce evolution is generally determined by mass accretion onto the bounce shock, which stalls shortly after its formation. The accumulated material settles onto the PNS surface to form a thick low-density layer. Of fundamental relevance of the PNS evolution, in particular the contraction behavior, is the EOS and the density dependence of associated nuclear matter properties. We have found that while the symmetry energy plays the dominant role during the core collapse phase until core bounce, it's role reduces slowly during the post bounce evolution. Instead, the symmetric part of the baryon EOS, e.g., free energy and pressure starts to dominate first at low densities and later at all densities. It is related to the continuously increasing temperatures obtained during the post bounce evolution prior to the possible onset of the explosion. Only close to and above saturation density, symmetric and asymmetric contributions become of similar magnitude, however, at these conditions the electron/positron 
EOS exceeds the baryon EOS. In this very aspect during the post bounce evolution, PNS and zero temperature as well as $\beta$-equilibrium neutron star physics differ substantially, since neutron star properties are dominantly determined by the symmetry energy, e.g., their radii.

Multi-dimensional simulations of neutrino-driven supernova explosions of massive stars have been discussed recently with regard to a comparison between the LS220 and STOS EOSs~\cite{Suwa:2013}. Here, we summarize results for the same low-mass 11.2~M$_\odot$ progenitor star as discussed above in spherical symmetry, comparing these two EOS. The softer LS EOS leads to more optimistic conditions for the explosion onset than the stiffer STOS. Note that in spherical symmetry the opposite holds. It is attributed to the larger mass enclosed inside the heating region as a direct structural feedback of the PNS of the multi-dimensional simulations, due to the presence of convection and the development of hydrodynamic instabilities. It becomes even more dramatic in case of a more massive 15~M$_\odot$ progenitor where neutrino driven explosions were obtained for LS but not for STOS~\cite{Suwa:2013}. The argument that neutrino-driven explosions are favored for a soft EOS has also been reported in ref.~\cite{Marek:2008qi}, 
applying in addition to LS an even softer EOS. Moreover, in the parametric study of ref.~\cite{Couch:2013} a similar conclusion has been achieved. However, any of these simulations were based on Newtonian physics and/or a simplified treatment of neutrino transport. It remains to be shown how much the conclusions may change when applying more advanced input physics, in particular general relativistic radiation hydrodynamics.

In addition to the standard supernova EOSs, we also discussed additional degrees of freedom which are not covered by saturation properties of nuclear matter at zero temperature. Therefore, we applied a new EOS that allows for the transition to strange quark matter above saturation density. It is based on the quark bag model and allows for massive neutron (hybrid) stars of about $2.0$~M$_\odot$. The appearance of strange quark matter at core bounce reduces the symmetry energy above saturation density. The associated softening of the high-density EOS results in higher central densities and lower electron fraction at core bounce. However, initial expectations about unstable PNS configurations that lead to a collapse, formation of a strong hydrodynamic shock, and subsequent explosion could not be fulfilled~\cite{Sagert:2008ka,Fischer:2011}. In addition to EOS uncertainties at high density, we also explored the presence of light nuclear clusters below saturation density. Although light clusters, such as deuteron 
and triton, can be abundant during the early post-bounce evolution prior to the explosion onset, their impact on the supernova dynamics via heating/cooling contributions from weak processes is expected to be small. In this article, we argue that it is because light clusters are only equally abundant as free protons in the region where the main part of the neutrino spectra is trapped. This aspect changes only after the explosion onset, when mass accretion vanishes and the PNS settles into a quasi-stationary state. Note that this analysis is based on the particular EOS HS(DD2). It may be altered when including a different nuclear interaction. During the subsequent PNS deleptonization, i.e. the Kelvin-Helmholtz cooling phase, the neutrino decoupling region shifts to higher densities where the light clusters can become even more abundant than free protons. Consequently, weak processes with light nuclear clusters may  impact the neutrino signal and the associated nucleosynthesis of heavy elements of the neutrino-
driven wind ejected from the PNS surface via continuous neutrino heating on a timescale of $10$ seconds. A further exploration of this important aspect requires the consistent inclusion of weak interaction rates with light nuclear clusters and corresponding EOS, as well as taking into account contributions from final-state Pauli blocking, in simulations that are based on accurate neutrino transport.

Moreover, the impact of the EOS during the PNS deleptonization phase can be very large. The thermodynamic properties (e.g. pressure and energy per baryon) of the EOS determine the PNS structure. In addition, different EOSs lead to a different nuclear composition which drives the deleptonization of the PNS via weak processes.  However, the neutrino luminosities and spectra which are obtained in long-term simulations of the PNS deleptonization show qualitative agreement for the two extreme EOS LS~\cite{Huedepohl:2010} and STOS~\cite{Fischer:2009af}. The aspect of potential convection inside the PNS during deleptonization and the possible impact of the symmetry energy has been explored recently~\cite{Roberts:2012f}. However, further explorations are required in order to obtain a systematic understanding of the impact of the symmetry energy on the PNS deleptonization as well as the subsequent mass ejection in the neutrino-driven wind.
%

%%%%%%%%%%%%%%%%%%%%%%%%%%%%%%%%%%%%%%%%%%%
\section*{Acknowledgment}
We thank Achim Schwenk for providing the Chiral EFT data shown in Fig.~\ref{fig:enm}. The spherically symmetric supernova simulations were performed at the computer cluster at the GSI Helmholtzzentrum f{\"u}r Schwerionenforschung GmbH, Darmstadt (Germany) and the axially symmetric simulations were in part carried on Cray XT4 and medium-scale clusters at CfCA of the National Astronomical Observatory of Japan, and on SR16000 at YITP in Kyoto University. TF acknowledges support from the Narodowe Centrum Nauki (NCN) within the "Maestro" program under contract No. DEC-2011/02/A/ST2/00306. MH is grateful for support from the Swiss National Science Foundation (SNF) under project number no. 200020-132816/1 and for participation in the ENSAR/THEXO project. I.S. is thankful to the Alexander von Humboldt foundation and acknowledges the support of the High Performance Computer Center and the Institute for Cyber-Enabled Research at Michigan State University. YS thanks the support by the Grants-in-Aid for the Scientific 
Research from the Ministry of Education, Culture, Sports, Science and Technology (MEXT), Japan (Nos. 23840023 and 25103511), and by HPCI Strategic Program of Japanese MEXT. The work of JSB was supported by the Hessian LOEWE initiative through the Helmholtz International Center for FAIR (HIC for FAIR) and the Alliance Program of the Helmholtz Association (HA216/EMMI).

%%%%%%%%%%%%%%%%%%%%%%%%%%%%%%%%%%%%%%%%%%%
\bibliography{manuscript}
\end{document}